\documentclass[sigconf]{acmart}

\usepackage{booktabs} 
\usepackage{multirow}
\usepackage{amsmath}
\usepackage{tabularx}
\usepackage{amsfonts}
\usepackage{arydshln}
\usepackage{subcaption}
\usepackage{microtype}
\usepackage[american]{babel}
\usepackage{enumitem}
\usepackage{algorithm}   
\usepackage{amssymb}
\usepackage[noend]{algcompatible}
\usepackage{stfloats} 
\fnbelowfloat

\def\b{{\bf b}}

\def\d{{\bf d}}

\def\p{{\bf p}}
\def\q{{\bf q}}

\def\w{{\bf w}}

\def\B{{\bf B}}

\def\D{{\bf D}}

\def\0{{\bf 0}}
\def\1{{\bf 1}}
\def\2{{\bf 2}}
\def\3{{\bf 3}}
\def\4{{\bf 4}}
\def\5{{\bf 5}}
\def\6{{\bf 6}}
\def\7{{\bf 7}}
\def\8{{\bf 8}}
\def\9{{\bf 9}}

\setcopyright{rightsretained}

\newcolumntype{R}{>{\raggedleft\arraybackslash}X}
\acmArticle{4}
\acmPrice{15.00}
\setcopyright{acmlicensed}
\begin{document}

\copyrightyear{2019} 
\acmYear{2019} 
\acmConference[CIKM '19]{The 28th ACM International Conference on Information and Knowledge Management}{November 3--7, 2019}{Beijing, China}
\acmBooktitle{The 28th ACM International Conference on Information and Knowledge Management (CIKM '19), November 3--7, 2019, Beijing, China}
\acmPrice{15.00}
\acmDOI{10.1145/3357384.3357930}
\acmISBN{978-1-4503-6976-3/19/11}



\title{Learning to Generate Candidates and Re-Rank for Large-Scale Top-N Recommendation}

\title{Candidate Generation with Binary Codes for Large-Scale Top-N Recommendation}



\author{Wang-Cheng Kang}
\affiliation{%
  \institution{University of California, San Diego}
  \city{La Jolla}
  \state{CA}
  \country{USA}
}
\email{wckang@ucsd.edu}

\author{Julian McAuley}
\affiliation{%
  \institution{University of California, San Diego}
  \city{La Jolla}
  \state{CA}
  \country{USA}
}
\email{jmcauley@ucsd.edu}

\renewcommand{\shortauthors}{Wang-Cheng Kang and Julian McAuley}

\begin{abstract}
Generating the Top-N recommendations
from a large corpus is 
computationally expensive to perform at scale.
Candidate generation and re-ranking based approaches are often adopted in industrial settings 
to alleviate efficiency problems. 
However it remains to be fully studied how well such schemes approximate complete rankings (or how many candidates are required to achieve a good approximation), or to develop systematic approaches to generate high-quality candidates efficiently.
In this paper, we seek to investigate these questions via proposing a
\underline{c}and\underline{i}date \underline{g}eneration \underline{a}nd \underline{r}e-ranking based framework (CIGAR), which first learns a preference-preserving binary embedding for building a hash table to retrieve candidates, and then learns to re-rank the candidates using real-valued ranking models with a candidate-oriented objective. 
We perform a comprehensive study on several large-scale real-world datasets 
consisting of
millions of users/items and hundreds of millions of interactions. Our results show that CIGAR significantly boosts the Top-N accuracy against state-of-the-art recommendation models, while reducing the query time by orders of magnitude. We hope that this work could draw more attention to the candidate generation problem in recommender systems.
\end{abstract}



\maketitle

\begin{figure}[t]
\centering
\includegraphics[width=\linewidth]{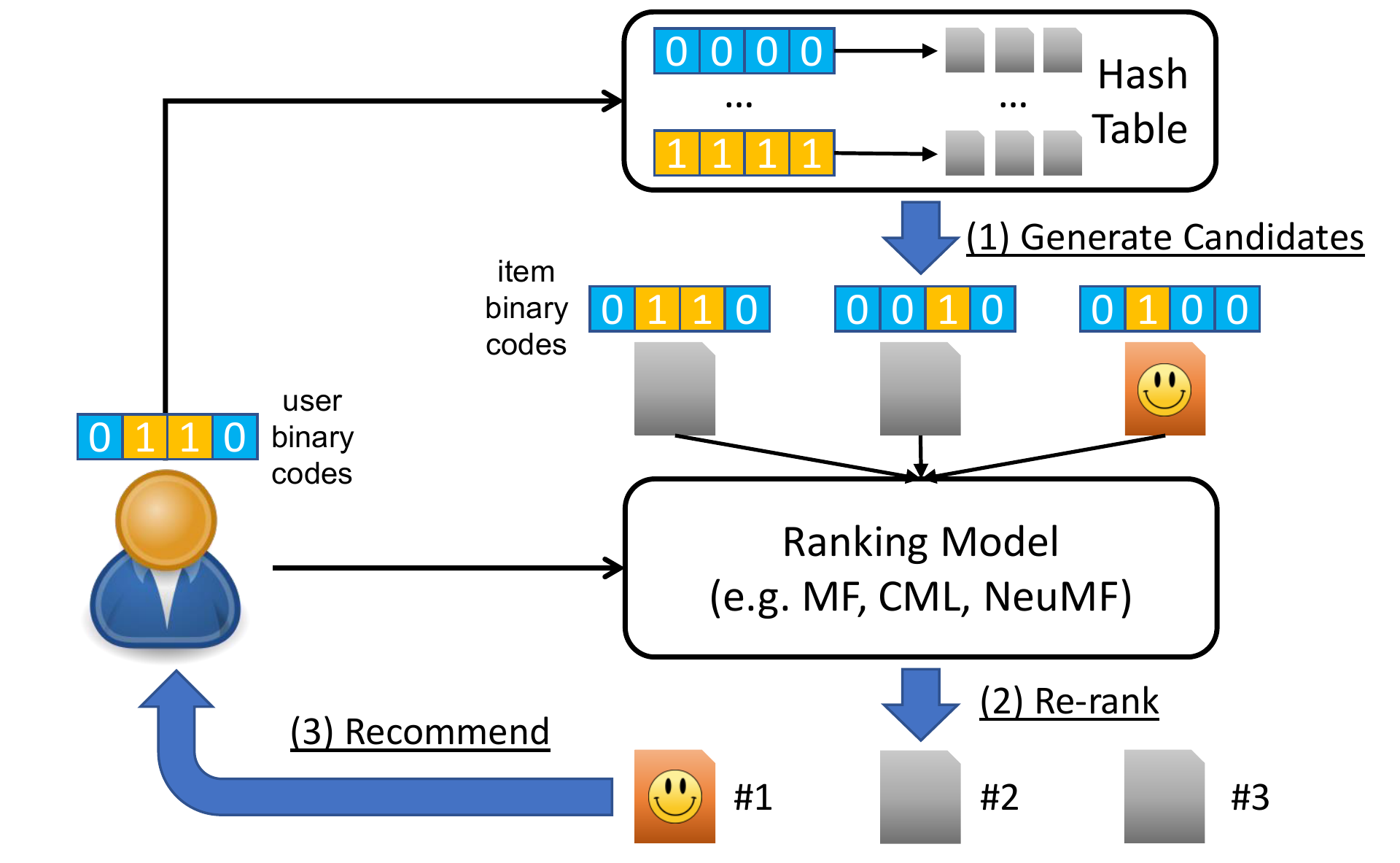}
\vspace{-0.4cm}
\caption{A simplified illustration showing the candidate generation and re-ranking procedures in our CIGAR framework. The binary codes and ranking model are both learned from user feedback.\label{fig:intro}}
\vspace{-0.4cm}
\end{figure}
\section{Introduction}

Top-N recommendation is a fundamental task of a recommender system, which consists of generating a (short) list of N items that are highly likely to be interacted with (e.g.~purchased, liked, etc.) by users. 
Precisely identifying these Top-N items from a large corpus is highly challenging, both from an accuracy and efficiency perspective. The vast number of items, 
both in terms of their
variability and sparsity, 
makes the problem especially difficult when scaling up to real-world datasets.
In particular,
exhaustively searching through all items to generate the Top-N ranking becomes intractable at scale due to its high latency.

Recommender systems have received significant attention
with various models being proposed, 
though generally focused on the goal of achieving better accuracy~\cite{DBLP:conf/wsdm/TangW18,DBLP:conf/kdd/Christakopoulou18a,hu2018leveraging,DBLP:conf/www/TayTH18,DBLP:journals/tois/DeshpandeK04}. For example, 
BPR-MF~\cite{rendle2009bpr} adopts a conventional Matrix Factorization~(MF) approach as its underlying preference model, CML~\cite{DBLP:conf/www/HsiehYCLBE17} employs metric embeddings,
TransRec~\cite{DBLP:conf/recsys/HeKM17} adopts translating vectors, and NeuMF~\cite{NeuMF} uses multi-layer perceptrons~(MLP) to model user-item interactions.

As for the problem of latency/efficiency, a few works seek to accelerate the maximum inner product~(MIP) search step (for MF-based models), via pruning or tree-based data structures~\cite{DBLP:conf/kdd/RamP12,DBLP:conf/sigmod/LiCYM17}. Such approaches are usually model-specific (e.g.~they depend on the specific structure of an inner-product space), 
and thus
are hard to generalize when trying to accelerate other models. Another line of work seeks to directly learn binary codes to estimate user-item interactions, and builds hash tables to accelerate retrieval time~\cite{DBLP:conf/sigir/ZhangSLHLC16,DBLP:conf/aaai/ZhangLY17,DBLP:conf/wsdm/ZhangYHDYL18,DBLP:conf/kdd/ZhangWLTYY18,DBLP:conf/kdd/LianLG00C17,DBLP:conf/ijcai/Liu0FNLZ18}. While using binary codes can significantly reduce query time to constant or sublinear complexity, the accuracy of such models is still inferior to conventional 
(i.e.,~real-valued) models, as such models are highly constrained,
and may lack sufficient flexibility when aiming to
precisely rank the Top-N items.

As 
the vast majority of items will be
irrelevant to most users at any given moment, candidate generation and re-ranking 
strategies
have been adopted in industry
where high efficiency is required.
Such approaches first generate a small number of candidates in an efficient way, and then apply fine-grained re-ranking methods to obtain the final ranking. To achieve high efficiency, the candidate generation stage is often based on rules or heuristics.
For example, \emph{Youtube's} early recommender system treated users' recent actions as seeds, and searched among relevant videos in the co-visitation graphs with a heuristic relevance score~\cite{DBLP:conf/recsys/DavidsonLLNVGGHLLS10}. \emph{Pinterest} performs random walks 
(again using recent actions as seeds) 
on the pin-board graph to retrieve relevant candidates~\cite{DBLP:conf/www/EksombatchaiJLL18}, and also considers other candidate sources based on various signals like annotations, content, etc.~\cite{DBLP:conf/www/LiuRSKMZLJ17}. Recently, \emph{Youtube} adopted deep neural networks~(DNNs) to extract user embeddings from various features, and used 
an inner product function to estimate scores,
such that
candidate generation can be accelerated by maximum inner product search~\cite{DBLP:conf/recsys/CovingtonAS16}.



In this paper, we propose a novel \underline{c}and\underline{i}date \underline{g}eneration \underline{a}nd \underline{r}e-ranking based framework called \emph{CIGAR}. During the candidate generation stage, unlike existing work that adopts heuristics, or learns real-valued embeddings first and then adopts indexing techniques to accelerate, we propose to directly learn binary codes for both preference ranking and hash table lookup. During the re-ranking stage, we learn to re-rank candidates using existing ranking models with candidate-oriented sampling strategies. Figure~\ref{fig:intro} shows the procedure of generating recommendations using CIGAR.

Our main contributions are 
as follows:
\begin{itemize}[leftmargin=5mm]
    \item We propose a novel framework (\textbf{CIGAR}) which learns to generate candidates with binary codes, and re-ranks candidates with real-valued models. CIGAR thus exhibits 
    both the efficiency of hashing and the accuracy of real-valued methods:
    binary codes are employed to estimate \emph{coarse-grained} preference scores and efficiently retrieve candidates, while real-valued models are used for \emph{fine-grained} re-ranking of a small number of candidates.
    \item We propose a new hashing-based method---\textbf{HashRec}---for learning binary codes with implicit feedback. HashRec is optimized via stochastic gradient descent, and can easily scale to large datasets. CIGAR adopts HashRec for fast candidate generation, as empirical results show that HashRec achieves superior performance compared to other hashing-based methods.
    \item We propose a candidate-oriented sampling strategy which encourages the models to focus on re-ranking candidates, rather than treating all items equally. With such a sampling scheme, CIGAR can significantly boost the accuracy of various exiting ranking models, including neural-based approaches. 
    \item Comprehensive experiments are conducted on several large-scale datasets. We find that CIGAR outperforms the existing state-of-the-art models, including those that rank all items, while reducing the query time by orders of magnitude. Our results suggest that it is possible to achieve similar or better performance 
    than existing approaches even when using
    only a small number of candidates. 
\end{itemize}

\section{Background}

In this section, we briefly review relevant background including representative ranking models for implicit feedback, and hashing-based models for efficient recommendation.

\subsection{Preference Ranking Models}
\subsubsection{Recommendation with Implicit Feedback}

In our paper, we focus on learning user preferences from implicit feedback (e.g.~clicks, purchases, etc.). Specifically, we are given a user set $\mathcal{U}$ and an item set $\mathcal{I}$, such that the set $\mathcal{I}_u^+$ represents the items that user $u$ has interacted with, while $\mathcal{I}_u^-=\mathcal{I}-\mathcal{I}_u^+$ represents unobserved 
interactions.
Unobserved interactions are not necessarily negative, 
rather for the majority of such items the user may simply be unaware of them.
To interpret such in-actions, weighted loss~\cite{WRMF} and learning-to-rank~\cite{rendle2009bpr} approaches have been proposed.  

\subsubsection{Bayesian Personalized Ranking (BPR)}

BPR~\cite{rendle2009bpr} is a classic approach for learning preference ranking models from implicit feedback. The core idea is to rank observed actions (items) higher than unobserved items. BPR-MF is a popular variant that adopts conventional Matrix Factorization (MF) approaches as its underlying preference estimator:
\begin{equation}\label{eq:bpr}
    s_{u,i} = \langle\p_u, \q_i\rangle,
\end{equation}
where $\p_u, \q_i$ are $k$-dimensional embeddings.
BPR seeks to optimize pairwise rankings by minimizing a contrastive objective:
\begin{equation}
\begin{split}
    -&\sum_{(u,i,j)\in \mathcal{D}}\ln\sigma (s_{u,i} - s_{u,j})\\
    \mathcal{D}=&\{(u,i,j)|u\in\mathcal{U}\land i\in\mathcal{I}^+\land j\in\mathcal{I}^-\}.
\end{split}
\end{equation}
As enumerating all triplets in $\mathcal{D}$ is typically intractable, BPR-MF adopts stochastic gradient descent~(SGD) to optimize the model. Namely, in each step of SGD, we dynamically sample a batch of triplets from $\mathcal{D}$. Also, an $\ell_2$ regularization on user and item embeddings is adopted, which is crucial to alleviate overfitting.

\subsubsection{Collaborative Metric Learning (CML)} Conventional MF-based methods 
operate in inner product spaces, which
are flexible but can easily overfit.
To this end, CML~\cite{DBLP:conf/www/HsiehYCLBE17} imposes the triangle inequality constraint, by adopting metric embeddings to represent users and items. Here the preference score is estimated by the 
negative $\ell_2$ distance:
\begin{equation}\label{eq:CML}
    s_{u,i} = -\|\p_u - \q_i\|_2.
\end{equation}
CML adopts a hinge loss to optimize pairwise rankings. A significant benefit of CML is that retrieval can be accelerated by efficient nearest neighbor search, which has been heavily studied.

\subsubsection{Neural Matrix Factorization~(NeuMF)} To estimate more complex and non-linear preference scores,
NeuMF~\cite{NeuMF} adopts multi-layer perceptrons for modeling interactions:
\begin{equation}\label{eq:NeuMF}
    s_{u,i} = \w^T \left[\begin{array}{c} \p^{(1)}_u\odot\q^{(1)}_i \\ \text{MLP}(\p^{(2)}_u, \q^{(2)}_i)\end{array}\right],
\end{equation}
where $\odot$ is the element-wise product, `MLP' extracts a vector from user and item embeddings, and $\w$ is used to project the concatenated vector to the final score. Essentially NeuMF combines generalized matrix factorization~(GMF) and MLPs. Due to the complexity of the scoring function, the retrieval process is generally hard to accelerate for NeuMF.

\subsection{Hashing-based Recommendation}
To achieve efficient recommendation, various hashing-based models have been proposed. These methods use binary representations to represent users and items, and the retrieval time can be reduced to constant or sublinear time by appropriate use of a hash table. We briefly introduce the Hamming Space and two relevant hashing-based recommendation method.
\subsubsection{Hamming Space} A Hamming space contains $2^r$ binary strings with length $r$. Binary codes can be efficiently stored and computed in modern 
systems. In this paper we use binary codes $\b_u,\d_i\in\{-1,1\}^r$ to represent users and items.\footnote{We use \{-1,1\} instead of \{0,1\} for convenience of formulations, though in practice we can convert to binary codes (i.e.,~\{0,1\}) when storing them.} The 
negative
Hamming distance measures the similarity between two binary strings:
\begin{equation}
\begin{split}
    s_{H}(\b_u, \d_i) &= \sum_{z=1}^{r} \mathbb{I}(b_{u,z}= d_{i,z})\\&=\frac{1}{2}\left(\sum_{z=1}^{r} \mathbb{I}(b_{u,z}= d_{i,z})+ r - \sum_{z=1}^{r} \mathbb{I}(b_{u,z}\neq d_{i,z})\right)
    \\&=\mathit{const} + \frac{1}{2}\langle\b_u, \d_i\rangle,
\end{split}
\label{eq:hamming}
\end{equation}
where $\mathbb{I}(\cdot)$ is the indicator function. This provides a convenient way to formulate the problem with the inner product.
\subsubsection{Discrete Collaborative Filtering~(DCF)} 
DCF~\cite{DBLP:conf/sigir/ZhangSLHLC16} is a representative method that estimates observed ratings (scaled to $[-r,r]$) using $\langle\b_u, \d_i\rangle$. Additional constraints of bit balance and bit uncorrelation are adopted to learn efficient binary codes. DCF introduces real-valued auxiliary variables, and adopts an optimization strategy
consisting of alternating sub-problems with closed-form solutions.

\subsubsection{Discrete Personalized Ranking~(DPR)} 
To our knowledge, DPR~\cite{DBLP:conf/aaai/ZhangLY17} is the only hashing-based method designed for implicit feedback. DPR considers triplets $\mathcal{D}$ as in BPR, and optimizes rankings using a squared loss. DPR also optimizes sub-problems with closed-form solutions. However, the solutions to these sub-problems rely on computing all triplets in $\mathcal{D}$, which 
makes  optimization hard to scale to large datasets.

\begin{table}[t]
\small
\caption{Notation. \label{tb:notation}}
\vspace{-0.2cm}
\begin{tabularx}{\linewidth}{lX}
\toprule
Notation&Description\\
\midrule
$\mathcal{U},\mathcal{I}$              & user and item set\\
$r\in \mathbb{N}$                      & binary embedding length (\#bits)\\
$k\in \mathbb{N}$                      & real-valued embedding size\\
$c\in \mathbb{N}$                      & number of candidates for re-ranking\\
$\widetilde\b_u, \widetilde\d_i\in\mathbb{R}^{r}$            & auxiliary embeddings for user $u$ and item $i$\\
$\b_u, \d_i\in\{-1,1\}^{r}$            & binary embeddings for user $u$ and item $i$\\
$\p_u, \q_i\in\mathbb{R}^{k}$          & real embeddings for user $u$ and item $i$\\
$m\in \mathbb{N}$                      & the number of substrings in MIH\\
$h\in \mathbb{R}$                      & the sampling ratio in eq.~\ref{eq:Dplus}\\
\bottomrule
\end{tabularx}
\end{table}

\section{CIGAR: Learning to Generate Candidates and Re-Rank}

In this section, we introduce CIGAR, a candidate generation and re-ranking based framework. We propose a new method HashRec that learns binary embeddings for users and items. CIGAR leverages the binary codes generated by HashRec, to construct hash tables for fast candidate generation. Finally, CIGAR learns to re-rank candidates via real-valued ranking models with the proposed sampling strategy. Our notation is summarised in Table~\ref{tb:notation}.

\begin{figure}[t]
\centering\small

\begin{subfigure}[t]{0.32\linewidth}
\includegraphics[width=\linewidth]{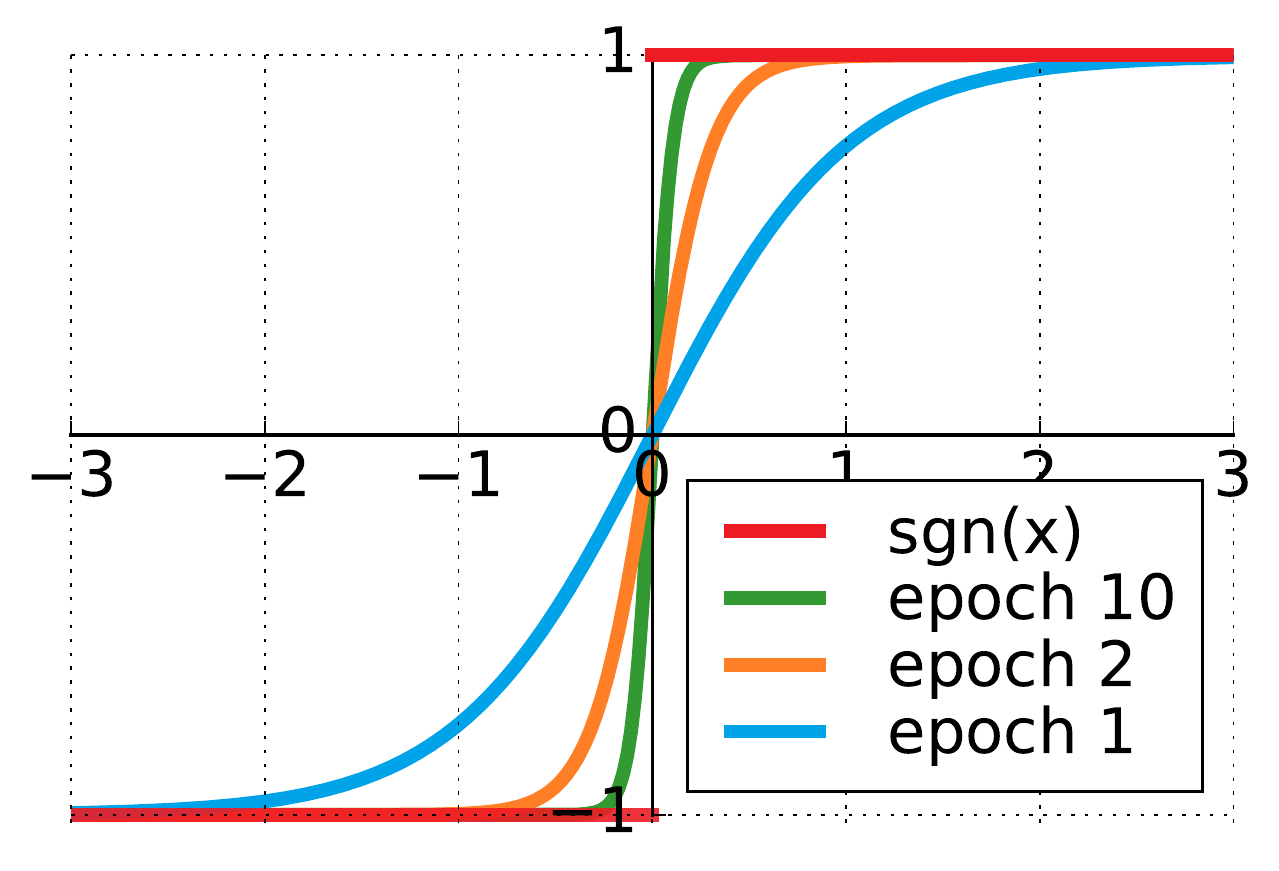}
\subcaption{\textmd{sgn}($x$) vs. $\tanh(\beta x)$}
\end{subfigure}
\begin{subfigure}[t]{0.32\linewidth}
\includegraphics[width=\linewidth]{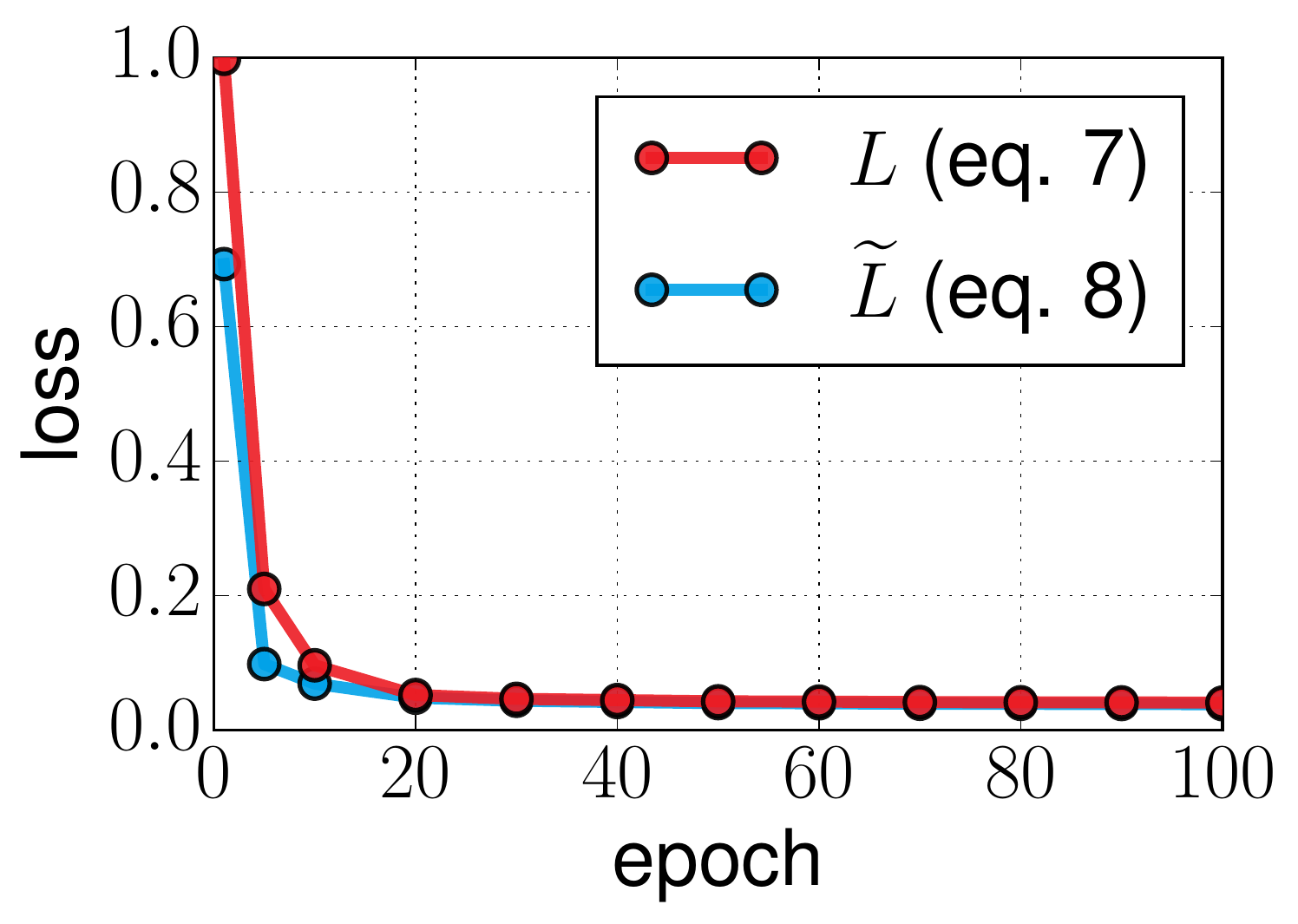}
\subcaption{Training loss}
\end{subfigure}
\begin{subfigure}[t]{0.32\linewidth}
\includegraphics[width=\linewidth]{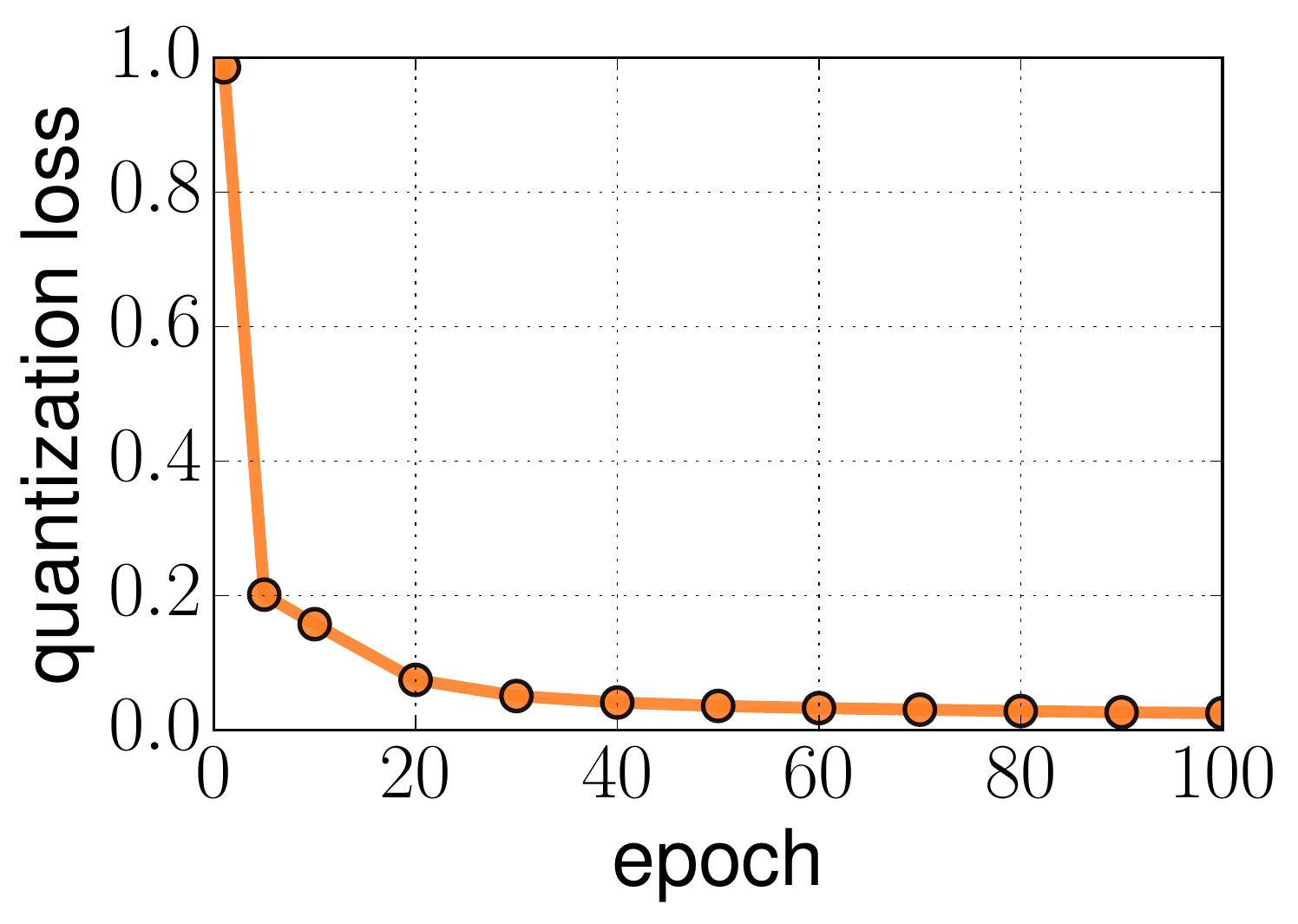}
\subcaption{Quantization loss}
\end{subfigure}
\vspace{-0.2cm}
\caption{Training curves on MovieLens-20M. Figure (a) plots \textmd{$\text{sgn}(x)$} and its approximation $\tanh(\beta x)$. Figure (b) plots the desired loss $L$ and surrogate loss $\widetilde{L}$ through training. Figure (c) shows the quantization error (measured via mean squared distances) between \textmd{$\text{sgn}(x)$} and $\tanh(\beta x)$.}
\label{fig:curves}
\end{figure} 

\subsection{Learning Preference-preserving Binary Codes}

We use binary codes $\b_u, \d_i\in\{-1, 1\}^r$ to represent users and items, and estimate interactions between them via the Hamming distance. We seek to learn preference-preserving binary codes such that similar binary codes (i.e.,~low Hamming distance) indicate high preference scores. For convenience, we use the conventional inner product (the connection is shown in eq.~\ref{eq:hamming}) in our formulation:
\begin{equation}
    s_{u,i} = \langle\b_u, \d_i\rangle.
\end{equation}

In implicit feedback settings, we seek to rank observed interactions ($u$,$i$) higher than unobserved interactions. To achieve this, we employ the classic BPR~\cite{rendle2009bpr} loss to learn our binary codes. However, directly optimizing such binary codes is generally NP-Hard~\cite{DBLP:conf/nips/WeissTF08}. Hence, we introduce auxiliary real-valued embeddings $\widetilde\b_u,\widetilde\d_i\in\mathbb{R}^r$ as used by other learning-to-hash approaches~\cite{DBLP:conf/sigir/ZhangSLHLC16}. Thus our objective is equivalent to:
\begin{equation}
L=-    \sum_{(u,i,j)\in \mathcal{D}} \ln\sigma_\alpha\left(\langle\text{sgn}(\widetilde\b_u), \text{sgn}(\widetilde\d_i)\rangle-\langle\text{sgn}(\widetilde\b_u),\text{sgn}(\widetilde\d_j)\rangle\right),
\label{eq:loss}
\end{equation}
where $\text{sgn}(x)=1$ if $x\geq0$ ($-1$ otherwise), and $\sigma_\alpha(x) = \sigma(\alpha x) = 1/(1+\exp(-\alpha x))$. As the inner product between binary codes can be large (i.e., $\pm r$), we set $\alpha<1$ to reduce the saturation zone of the sigmoid function. Inspired by a recent study for image hashing~\cite{DBLP:conf/aaai/JiangL18,DBLP:conf/iccv/CaoLWY17}, we seek to optimize the problem by approximating the $\text{sgn}(\cdot)$ function:
\begin{equation*}
    \text{sgn}(x) = \lim_{\beta\to\infty}\tanh(\beta x),
\end{equation*}
where $\beta$ is a hyper-parameter that increases during training. With this approximation, the objective becomes:
\begin{equation}
\begin{split}
\widetilde{L}=-    \sum_{(u,i,j)\in \mathcal{D}} \ln\sigma_\alpha&\left(\langle\tanh(\beta\widetilde\b_u)\right., \tanh(\beta\widetilde\d_i)\rangle-\\ &\left.\langle\tanh(\beta\widetilde\b_u),\tanh(\beta\widetilde\d_j)\rangle\right).
\label{eq:loos_appr}
\end{split}
\end{equation}
As shown in Figure~\ref{fig:curves}, when we optimize the surrogate 
loss $\widetilde{L}$, the desired loss $L$ is also minimized consistently. 
Also we can see that the quantization loss (i.e., the mean squared distances between $\text{sgn}(x)$ and $\tanh(\beta x)$) drops significantly throughout the training process. Note that we also employ $\ell_2$-regularization on embeddings $\widetilde\b_u$ and $\widetilde\d_i$, as in BPR.

We name this method \textbf{HashRec}; the complete algorithm 
is given in Algorithm~\ref{algo:hasrec}.


\begin{algorithm}[t]
\small
\caption{Optimization in HashRec}
\label{algo:hasrec}
\begin{algorithmic}
\STATE {\textbf{Input:} training data $\mathcal{D}$, code length $r$, regularization coefficient $\lambda$}
\STATE {Initialize embeddings $\widetilde\B\in\mathbb{R}^{|\mathcal{U}|\times r}$ and $\widetilde\D\in\mathbb{R}^{|\mathcal{I}|\times r}$ (at random)}
\FOR{$\mathit{epoch} = 1 \to \mathit{num\_epochs}$}
  \STATE {$\beta \gets \sqrt{10*(\mathit{epoch}-1)}$}
  \FOR{$\mathit{iter} = 1 \to \mathit{num\_iterations}$}
      \STATE{Sample a batch of triplets a from $\mathcal{D}$}
      \STATE{Optimize loss~(\ref{eq:loos_appr}) by updating $\widetilde\b_u$ and $\widetilde\d_i$ using the Adam~\cite{DBLP:journals/corr/KingmaB14} optimizer}
  \ENDFOR
\ENDFOR
\STATE {Obtain binary codes by $\b_u\gets\text{sgn}(\widetilde\b_u)$ and $\d_i\gets\text{sgn}(\widetilde\d_i)$}
\STATE {\textbf{Output:} $\B\in\{-1,1\}^{|\mathcal{U}|\times r}$, $\D\in\{-1,1\}^{|\mathcal{I}|\times r}$}
\end{algorithmic}
\end{algorithm}

\subsection{Building Multi-Index Hash Tables}

Using binary codes to represent users and items can yield significant benefits in terms of storage cost and retrieval speed. For example, in our experiments, HashRec achieves satisfactory accuracy with $r$=64 bits, which is equivalent in space to only 4 single-precision floating-point numbers (i.e.,~float16). Moreover,
computing architectures are amenable to calculating the Hamming distance of binary codes.\footnote{The Hamming distance can be efficiently calculated by two instructions: XOR and POPCNT (count the number of bits set to 1).} In other words, performing exhaustive search with binary codes is much faster (albeit by a constant factor) compared to real-valued embeddings. However, using exhaustive search inevitably leads to linear time complexity (in $|\mathcal{I}|$), which still scales poorly.

To scale to large, real-world datasets, we seek to build hash tables to index all items according to their binary codes, such that we can perform hash table lookup to retrieve and recommend items for a given user. Specifically, for a query code $\b_u$, we retrieve items from buckets within a small radius $l$ (i.e., $d_H(\b_u,\d_i)\leq l$). Hence the returned items have low Hamming distances (i.e., high preference scores) compared to the query codes, and 
search can be done in constant time.
However, for large code lengths the number of buckets grows exponentially, and furthermore such an approach may return zero items as nearby buckets 
will frequently be empty due to dataset sparsity.

Hence we employ Multi-Index Hashing (MIH)~\cite{DBLP:journals/pami/0002PF14} as our indexing data structure. The core idea of MIH is to split binary codes to $m$ substrings and index them by $m$ hash tables.
When we retrieve items within Hamming radius $l$, we first retrieve items in each hash table with radius $\lfloor\frac{l}{m}\rfloor$, and then sort the retrieved items based on their Hamming distances with the full binary codes. It can be guaranteed that such an approach can retrieve the desired items (i.e., within Hamming radius $l$), and that the search time is sub-linear in the number of items~\cite{DBLP:journals/pami/0002PF14}. 

Since we are interested in generating fixed-length (Top-N) rankings,
we seek to retrieve $c$ items as candidates, instead of considering Hamming radii. MIH proposes an adaptive solution that gradually increases the radius $l$ until enough items are retrieved (i.e., at least $c$). Empirically we found that the query time of MIH is extremely fast and grows slowly with the number of items. The pseudo-code for constructing and retrieving items in MIH, and more information about this process is described in the 
appendix.

\subsection{Candidate-oriented Re-ranking}\label{sec:COS}

So far we have
learned preference-preserving binary codes for users and items, and constructed hash tables to efficiently retrieve items for users. However, as observed in previous hashing-based methods, generating recommendations purely using binary codes leads to inferior accuracy compared with conventional real-valued ranking models. To achieve satisfactory performance in terms of both accuracy and efficiency, we propose to use the retrieved items as candidates, and adopt sophisticated ranking models to refine the results. As the preference ranking problem has been heavily studied~\cite{rendle2009bpr,DBLP:conf/www/HsiehYCLBE17,NeuMF}, we employ existing models to study the effect of the CIGAR framework, and propose a candidate-oriented sampling strategy to further boost accuracy.

A straightforward approach would be to adopt `off-the-shelf' ranking models (e.g.~BPR-MF) for re-ranking. However, we argue that such an approach is sub-optimal as existing models are typically trained 
to produce
rankings for all items, while our re-ranking models only rank the $c$ generated candidates. Moreover, the retrieved candidates are often `difficult' items (i.e., items that are hard for ranking models to discriminate) 
or at the very least are not a uniform sample of items.
Hence, 
it might be
better to train ranking models such that they are focused on the re-ranking objective. In this section, we introduce our candidate-oriented sampling strategy, and show how to apply it to existing ranking models in general. 

The loss functions of preference ranking models can generally be described as\footnote{For pairwise learning based methods (e.g.~NeuMF), we have $\mathcal{L}(u,i,j)=\mathcal{L}^+(u,i)+\mathcal{L}^-(u,j)$.}:
\begin{equation}
\min \sum_{(u,i,j)\in \mathcal{D}} \mathcal{L} (u,i,j).
\end{equation}
The choice of $\mathcal{L}$ is flexible, for instance, BPR uses $\ln\sigma(\cdot)$, and CML adopts the margin loss. To make the model focus more on learning to re-rank the candidates, we propose a candidate-oriented sampling strategy, which substitutes $\mathcal{D}$ with
\begin{equation}\label{eq:Dplus}
\mathcal{D}^+ = \left\{\begin{array}{ll}
\text{sample from }\mathcal{D}, & \text{with probability } 1-h\\
\text{sample from }\mathcal{C}, & \text{with probability } h
\end{array}\right.,
\end{equation}
where $\mathcal{C}$=$\{(u,i,j)|u\in\mathcal{U}\land i\in\mathcal{I}_u^+\land j\in\mathcal{I}_u^-\cap\mathcal{C}_u\}$, $\mathcal{C}_u$ contains $c$ candidates for user $u$ generated by hashing,
and $h$ controls the 
probability
ratio. Note that the sampling is equivalent to assigning larger weights to the candidates (for $h$>0). 
We empirically find that the best performance is obtained with 0<$h$<1;
when $h$=0, the model is not aware of the candidates that need to be ranked, while $h$=1 may lead to overfitting due to the limited number of samples. 

As for constructing $\mathcal{C}$, one approach is online generation, as we did for $\mathcal{D}$. Namely, in each step of SGD, we sample a batch of users and obtain candidates by hashtable lookup. Another approach is to pre-compute and store all candidates. Both approaches are practical, though we adopt the latter for better time efficiency.

Finally, taking BPR-MF as an example, the candidate-oriented re-ranking model is trained with the following objective:
\begin{equation}
 -\sum_{(u,i,j)\in \mathcal{D}^+} \ln\sigma \left(\langle\p_u, \q_i\rangle-\langle\p_u, \q_j\rangle\right).
\end{equation}
We denote this model as BPR-MF\textsuperscript{+}, to distinguish against the vanilla model. CML\textsuperscript{+} and NeuMF\textsuperscript{+} (etc.)~are denoted in the same way.

\subsection{Summary}

To summarize, the training process of CIGAR consists of: (1) Learning preference-preserving binary codes $\b_u$ and $\d_i$ using HashRec~(Algorithm~\ref{algo:hasrec}); (2) Constructing Multi-Index Hash tables to index all items; (3) Training a ranking model (e.g.~BPR-MF\textsuperscript{+}) for re-ranking candidates (i.e., using a candidate-oriented sampling strategy).
We adopt SGD to learn binary codes as well as our re-ranking model, such that optimization easily scales to large datasets. 

During testing, for a given user $u$, we first retrieve $c$ candidates via hashtable lookup, then we adopt a linear scan to calculate their scores estimated by the re-ranking model, and finally return the Top-N items with the largest scores. Using candidates generated by hash tables significantly reduces the time complexity from linear (i.e.,~exhaustive search) to sub-linear, 
and in practice is over 1,000 times faster for 
the largest datasets in our experiments.

For hyper-parameter selection, by default, we set the number of candidates $c$=200, the number of bits $r$=64, the scaling factor $\alpha$=$10/r$, and sampling ratio $h$=0.5. The $\ell_2$ regularizer $\lambda$ 
is
tuned via cross-validation. In MIH, the number of substrings $m$ is manually set depending on dataset size. For example, we set $m$=4 for datasets with millions of items. Further details are included in the appendix. 




\section{Experiments}

We conduct comprehensive experiments to answer the following research questions:

\begin{description}
\item[\textbf{RQ1:}] Does CIGAR achieve similar or better Top-N accuracy compared with state-of-the-art models 
that perform exhaustive rankings (i.e., which rank all items)?
\item[\textbf{RQ2:}] Does CIGAR accelerate the retrieval time of inner-product-, metric-, or neural-based models on large-scale datasets?
\item[\textbf{RQ3:}] Does HashRec outperform alternative hashing-based approaches? 
Do candidate-oriented sampling strategies (e.g.~BPR-MF\textsuperscript{+}) help?
\item[\textbf{RQ4:}]  What is the influence of key hyper-parameters in CIGAR?
\end{description}

The code and data processing script are available at \url{https://github.com/kang205/CIGAR}.

\subsection{Datasets}
We evaluate our methods on four public benchmark datasets. A proprietary dataset from \emph{Flipkart}
is also employed to test the scalability of our approach on a representative industrial dataset.
The datasets vary significantly in domains, sparsity, and number of 
users/items;
dataset statistics are shown in Table~\ref{tb:dataset}.
We consider the following datasets:
\begin{itemize}[leftmargin=5mm]
\item \textbf{MovieLens\footnote{\url{https://grouplens.org/datasets/movielens/20m/}}} A widely used benchmark dataset for evaluating collaborative filtering algorithms~\cite{DBLP:journals/tiis/HarperK16}. We use the largest version that includes 20 million user ratings. We treat all ratings as implicit feedback instances
(since we are trying to predict whether users will \emph{interact} with items, rather than their ratings).

\item \textbf{Amazon\footnote{\url{http://jmcauley.ucsd.edu/data/amazon/index.html}}} A series of datasets introduced in~\cite{VisualSIGIR}, including large corpora of product reviews crawled from \emph{Amazon.com}. Top-level product categories on \emph{Amazon} are treated as separate datasets, and we use the largest category `books.' All reviews 
are treated as implicit feedback.

\item \textbf{Yelp\footnote{\url{https://www.yelp.com/dataset/challenge}}} Released by \emph{Yelp}, containing various 
metadata about
businesses (e.g.~location, category, opening hours) as well as user reviews. We use the Round-12 version, and regard all review actions as implicit feedback.

\item \textbf{Goodreads.\footnote{\url{https://sites.google.com/eng.ucsd.edu/ucsdbookgraph/home}}} A recently introduced large dataset containing book metadata and user actions (e.g.~shelve, read, rate)~\cite{DBLP:conf/recsys/WanM18}. We treat the most abundant action (`shelve') as implicit feedback. As shown in~\cite{DBLP:conf/recsys/WanM18}, the dataset is dominated by a few popular items (e.g.~over 1/3 of users added \emph{Harry Potter \#1} to their shelves), such that always recommending the most popular books achieves high Top-N accuracy;
we ignore such outliers by
discarding the 0.1\% of most popular books.

\item \textbf{Flipkart} A large dataset of user sessions from \emph{Flipkart.com}, a large online electronics and fashion retailer in India. The recorded actions include `click,' `purchase,' and `add to wishlist.' Data was crawled over November 2018.
We treat all actions as implicit feedback.
\end{itemize}

For all datasets,
we take the k-core of the graph to ensure that all users and items have at least 5 interactions.
Following~\cite{rendle2009bpr,DBLP:conf/kdd/Christakopoulou18a}, we adopt
a
\emph{leave-one-out} strategy for data partitioning: for each user, we  randomly select two actions, put them into a validation set and test set respectively, and 
use all remaining actions
for training.

\begin{table}[t]
\centering\small
\caption{Dataset statistics (after preprocessing) \label{tb:dataset}}
\vspace{-0.2cm}
\begin{tabularx}{.95\linewidth}{Xcccc}
\toprule
Dataset         & \#Items   &   \#Users & \#Actions & \% Density\\ \midrule
MovieLens-20M   &   18K     &   138K    &   20M     &   0.81 \\
Yelp            &   103K    &   244K    &   3.7M    &   0.015 \\
Amazon Books    &   368K    &   604K    &   8.9M    &   0.004 \\
GoodReads       &   1.6M    &   759K    &   167M    &   0.014 \\
Flipkart            &   2.9M    &   9.0M      & 274M  &   0.001\\
\bottomrule
\end{tabularx}
\vspace{-0.4cm}
\end{table}

\subsection{Evaluation Protocol}

We adopt two common Top-N metrics: Hit Rate (HR) and Mean Reciprocal Rank (MRR), to evaluate recommendation performance~\cite{DBLP:conf/kdd/Christakopoulou18a,DBLP:conf/recsys/HeKM17}. HR@N counts the fraction of times that the single left-out item (i.e., the item in the test set) is ranked among the top N items, while MRR@N is a position-aware metric which assigns larger weights to higher positions (i.e., $1/i$ for the $i$-th position). Note that since we only have one test item for each user, HR@N is equivalent to Recall@N, and is proportional to Precision@N. Following~\cite{DBLP:conf/kdd/Christakopoulou18a}, we set N to 10 by default.

\begin{table*}
\centering\small
\caption{Recommendation performance. The best performing method in each row is boldfaced, and the second best method in each row is underlined. All the numbers are shown in percentage. - means the training fails due to lack of memory.
\label{tab:recommendation}}\vspace{-0.3cm}
\begin{tabular}{llcccc;{2pt/2pt}lcccccccc}
\toprule
\multirow{2}{*}{Dataset} & \multirow{2}{*}{Metric}     & \multirow{2}{*}{\begin{tabular}[c]{@{}c@{}}(a-1)\\ POP\end{tabular}} & \multirow{2}{*}{\begin{tabular}[c]{@{}c@{}}(a-2)\\ BPR-B\end{tabular}} & \multirow{2}{*}{\begin{tabular}[c]{@{}c@{}}(a-3)\\ DCF\end{tabular}} & \multirow{2}{*}{\begin{tabular}[c]{@{}c@{}}(a-4)\\ HashRec\end{tabular}} &
\multirow{2}{*}{Metric} &
\multirow{2}{*}{\begin{tabular}[c]{@{}c@{}}(b-1)\\ BPR-MF \end{tabular}}&
\multirow{2}{*}{\begin{tabular}[c]{@{}c@{}}(b-2)\\ CML \end{tabular}}& 
\multirow{2}{*}{\begin{tabular}[c]{@{}c@{}}(b-3)\\ NeuMF\end{tabular}}&
\multirow{2}{*}{\begin{tabular}[c]{@{}c@{}}\begin{scriptsize}CIGAR0\end{scriptsize}\\ \begin{scriptsize}HashRec+BPR-MF\end{scriptsize}\end{tabular}}&
\multirow{2}{*}{\begin{tabular}[c]{@{}c@{}}\begin{scriptsize}CIGAR\end{scriptsize}\\ \begin{scriptsize}HashRec+BPR-MF\textsuperscript{+}\end{scriptsize}\end{tabular}}&
\multirow{2}{*}{Improv. \%} \\ 
\\ \midrule
\multirow{2}{*}{\textbf{ML-20M}}    & HR@10  & 8.15 & 11.85 & 6.90 & \textbf{15.72}& HR@10 & 21.05 & \underline{21.41} & 18.43 &   21.56     & \textbf{25.42} & 18.7 \\
                                    & HR@200  & 42.35 & 51.88 & 43.31 & \textbf{62.96} &  MRR@10 & 8.82 & \underline{8.92} & 7.01 &  8.78      & \textbf{11.46} & 28.4  \\[1.5mm]
\multirow{2}{*}{\textbf{Yelp}}      & HR@10  & 1.03 & 0.21 & 1.30 & \textbf{1.39} &  HR@10 & \underline{2.64} & 2.12 & 1.70 &  2.82      & \textbf{3.33}& 26.1 \\
                                    & HR@200  & 7.82 & 3.38 & 10.19 & \textbf{18.49} &  MRR@10 & \underline{0.88} & 0.69 & 0.57 &   0.97     & \textbf{1.22}& 38.6  \\[1.5mm]
\multirow{2}{*}{\textbf{Amazon}}    & HR@10  & 0.74 & 1.26 & 0.91 & \textbf{2.08} &  HR@10 & \underline{4.17} & 3.47 & 1.59 &  3.73      & \textbf{4.56}& 9.4 \\
                                    & HR@200  & 5.14 & 6.56 & 6.45 & \textbf{11.69} &MRR@10 & \underline{1.73} & 1.41 & 0.67 &   1.91     & \textbf{2.23}& 28.9  \\[1.5mm]
\multirow{2}{*}{\textbf{GoodReads}} & HR@10  & 0.06 & 0.44 & 0.40 & \textbf{1.19} & HR@10 & 3.07 & \underline{3.20} & 2.20 &  2.46      & \textbf{3.39}& 5.9 \\
                                    & HR@200  & 1.61 & 2.86 & 3.99 & \textbf{8.10} & MRR@10 & \underline{1.42} & 1.26 & 0.91 &  1.15      & \textbf{1.73}& 21.8  \\[1.5mm]
\multirow{2}{*}{\textbf{Flipkart}}      & HR@10  & 0.05 & 0.34 & -  & \textbf{0.92}&HR@10 & \underline{2.68} & 0.81 & - &  1.68      & \textbf{2.74}& 2.2 \\
                                    & HR@200  & 1.24 & 1.92 & - & \textbf{6.13}& MRR@10 & \underline{1.03} & 0.30 & - &  0.64      & \textbf{1.19}& 13.6  \\                                
\bottomrule
\end{tabular}
\vspace{-0.1cm}
\end{table*}

\begin{figure*}
\centering\small
\begin{subfigure}[b]{\textwidth}
\centering
\includegraphics[width=.99\linewidth]{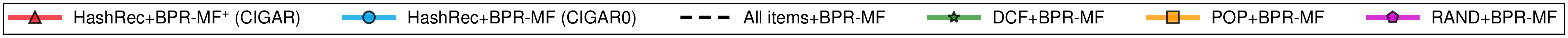}
\end{subfigure}
\begin{subfigure}[b]{0.195\textwidth}
\includegraphics[width=\linewidth]{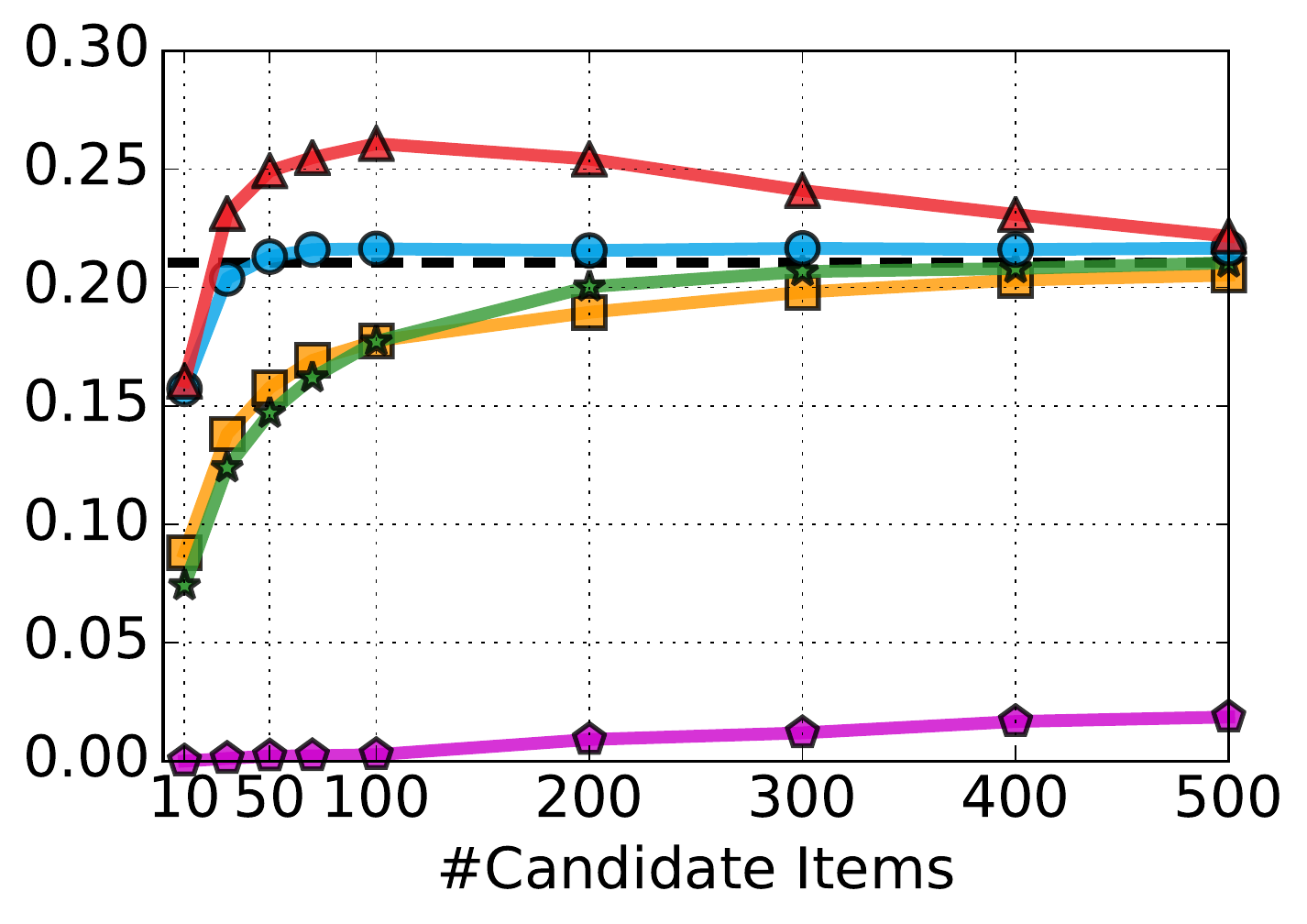}
\subcaption{ML-20M}
\end{subfigure}
\begin{subfigure}[b]{0.195\textwidth}
\includegraphics[width=\linewidth]{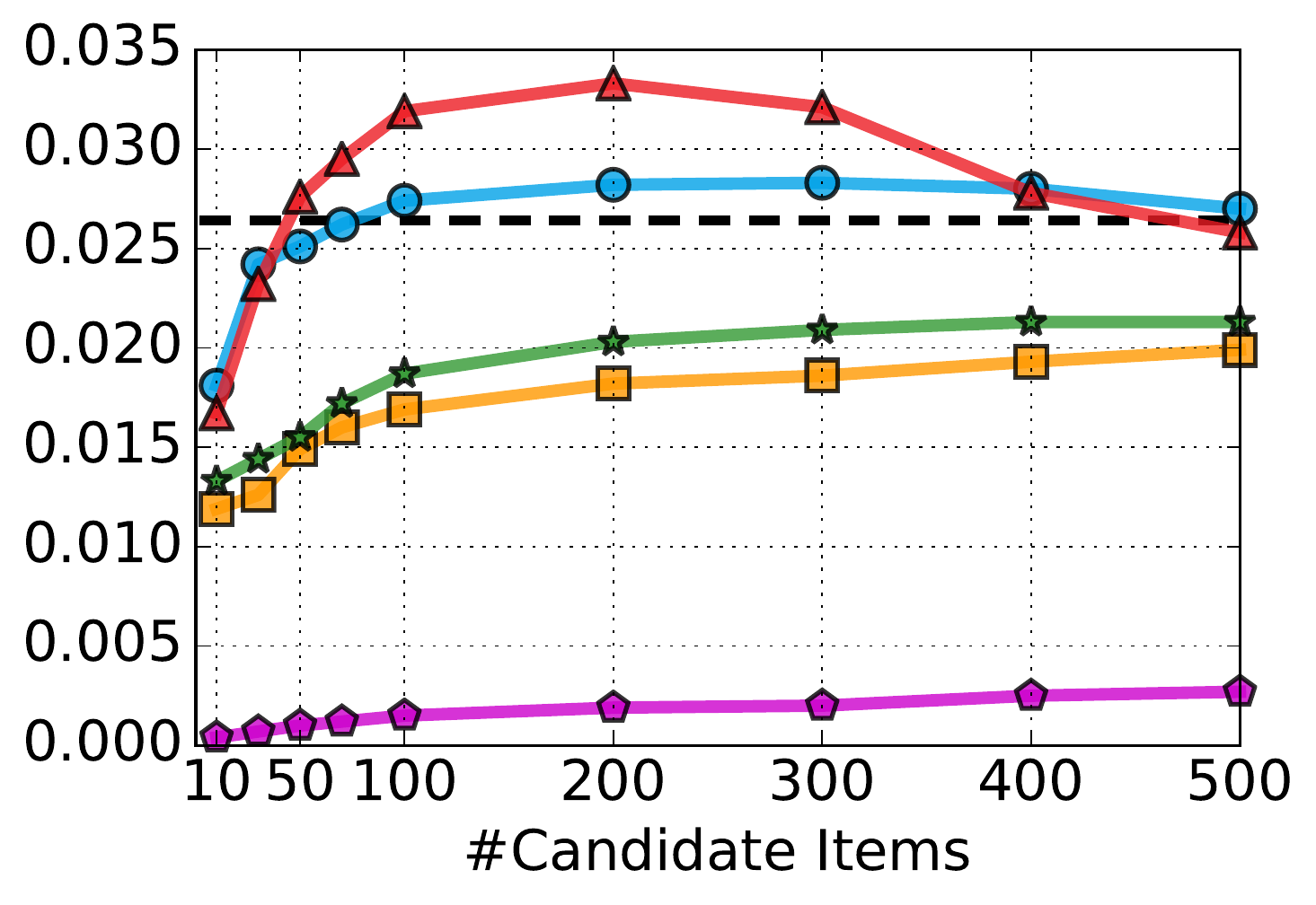}
\subcaption{Yelp}
\end{subfigure}
\begin{subfigure}[b]{0.195\textwidth}
\includegraphics[width=\linewidth]{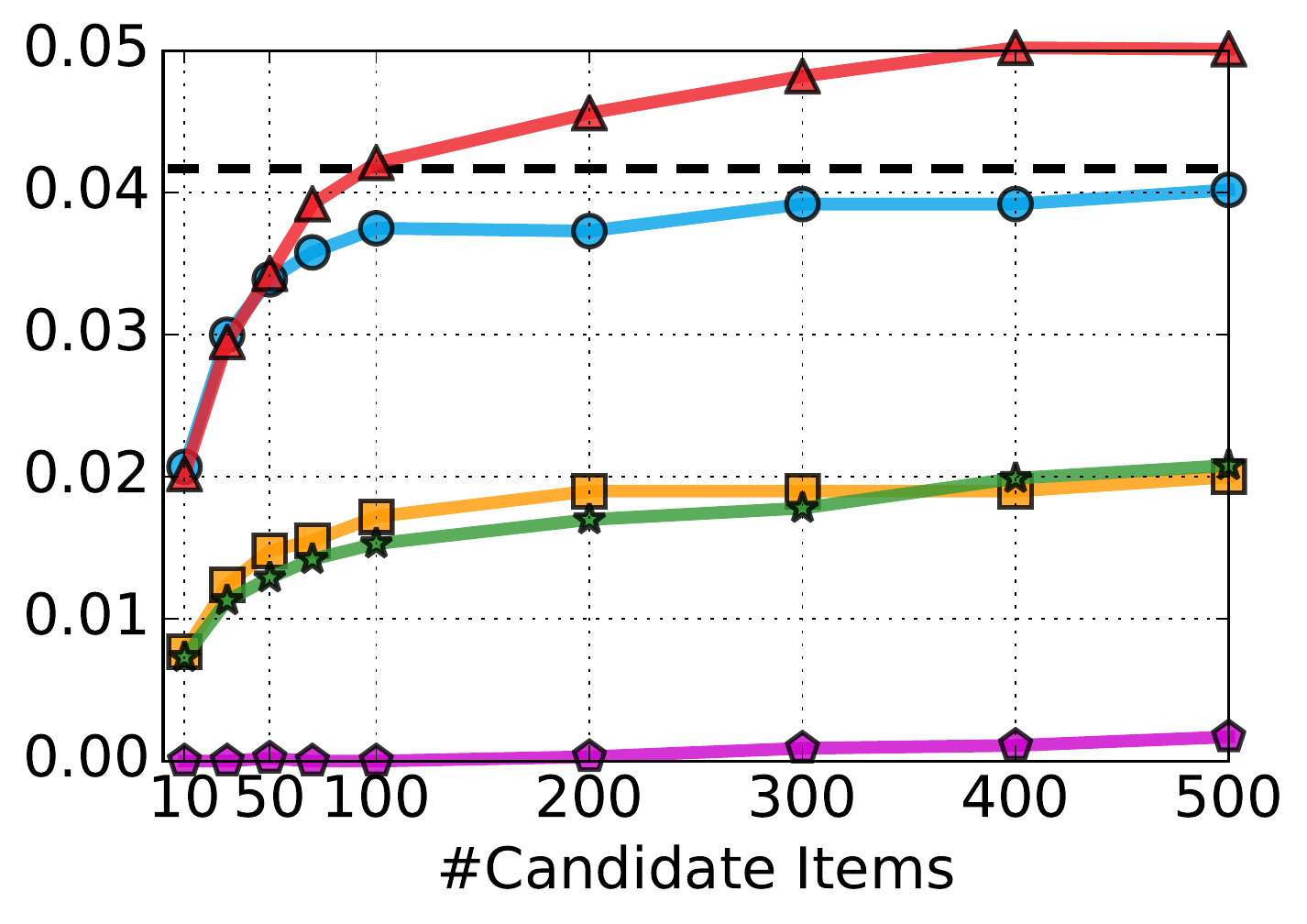}
\subcaption{Amazon Books}
\end{subfigure}
\begin{subfigure}[b]{0.195\textwidth}
\includegraphics[width=\linewidth]{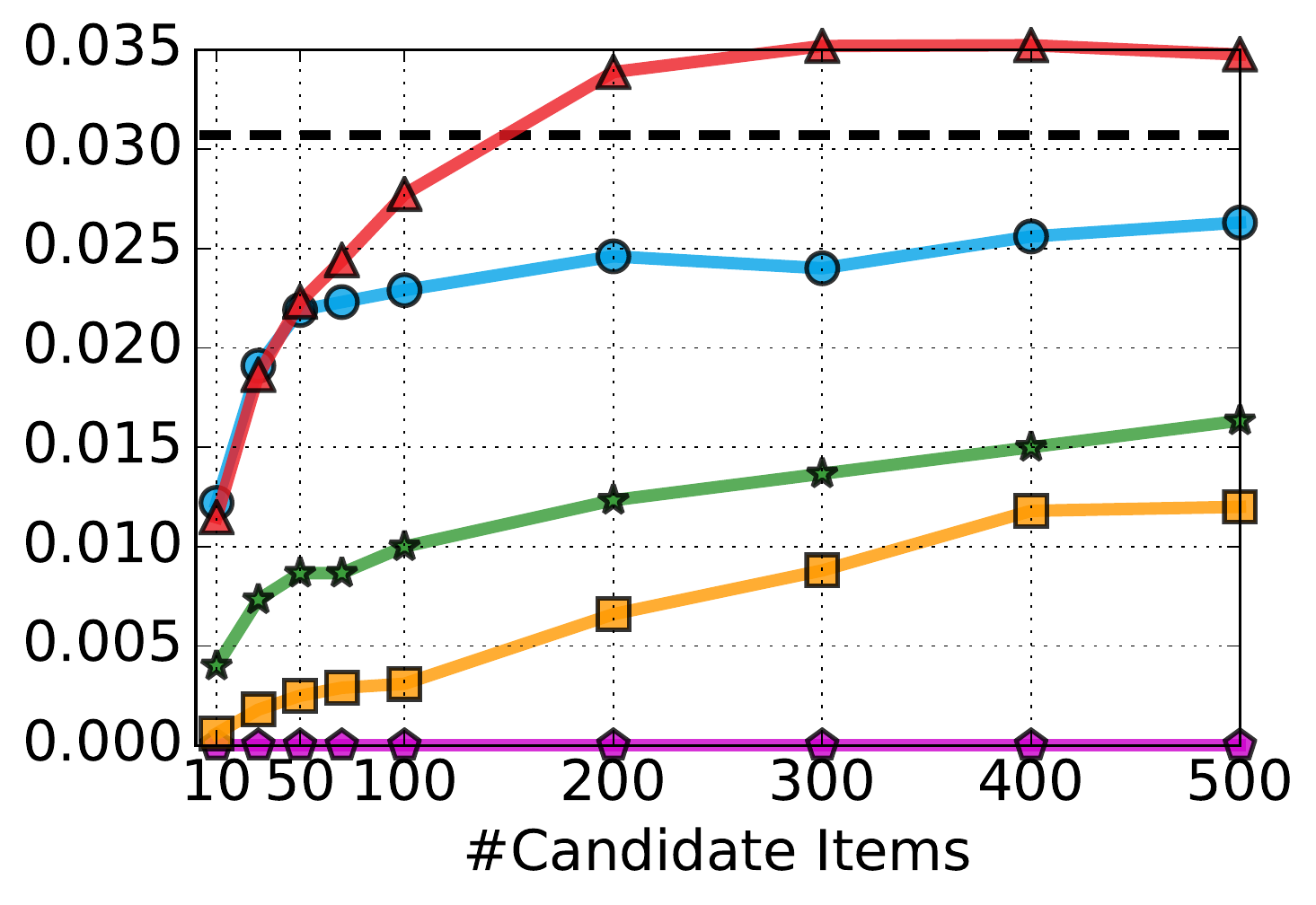}
\subcaption{Goodreads}
\end{subfigure}
\begin{subfigure}[b]{0.195\textwidth}
\includegraphics[width=\linewidth]{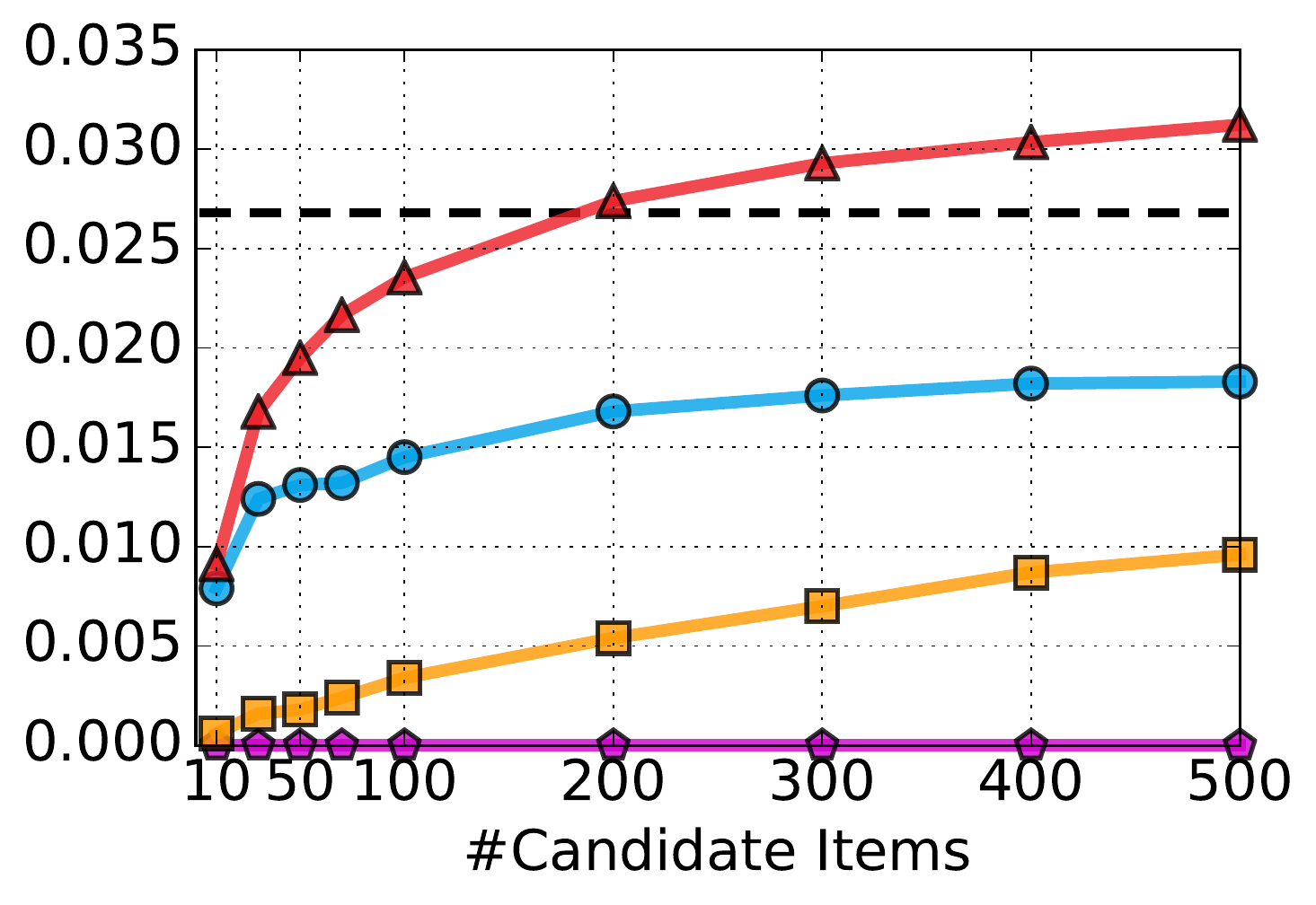}
\subcaption{Flipkart}
\end{subfigure}
\vspace{-0.2cm}
\caption{Effect of the number of candidates.}
\vspace{-0.2cm}
\label{fig:NC}
\end{figure*} 

\subsection{Baselines}

We consider three representative recommendation models that estimate user-item preference scores with inner-products, Euclidean distances, and neural networks:
\begin{itemize}[leftmargin=5mm]
\item \textbf{Bayesian Personalized Ranking (BPR-MF)~\cite{rendle2009bpr}} A classic model that seeks to optimize a pairwise ranking loss. We employ MF as its preference predictor as shown in eq.~\ref{eq:bpr}. When recommending items, a maximum inner product search is needed.
\item \textbf{Collaborative Metric Learning (CML)~\cite{DBLP:conf/www/HsiehYCLBE17}} CML represents users and items in a metric space, and measures their compatibility via the Euclidean distance (as shown in eq.~\ref{eq:CML}). The recommended items for a user can be retrieved via nearest neighbor search.
\item \textbf{Neural Matrix Factorization (NeuMF)~\cite{NeuMF}} NeuMF models non-linear interactions between user and item embeddings via a multi-layer perceptron (MLP). A generalized matrix factorization (GMF) term is also included in the model as in
eq.~\ref{eq:NeuMF}.
\end{itemize}

We also compare  HashRec with various hashing-based recommendation approaches:
\begin{itemize}[leftmargin=5mm]
\item \textbf{POP} A na\"ive popularity-oriented baseline
that simply ranks items by their global popularity.
\item \textbf{BPR-B} A simple baseline that directly 
quantizes embeddings from BPR-MF (i.e., applying $\text{sgn}(x)$ to the embeddings).
\item \textbf{Discrete Collaborative Filtering (DCF)~\cite{DBLP:conf/sigir/ZhangSLHLC16}} DCF learns binary embeddings to estimate observed ratings. To adapt it to the implicit feedback setting, we treat all actions 
as having value 1
and randomly sample 100 unobserved items (for each user) 
with value 0.
\item \textbf{Discrete Personalized Ranking (DPR)~\cite{DBLP:conf/aaai/ZhangLY17}} DPR is a hashing-based method designed for optimizing ranking with implicit feedback. However, due to its high training complexity, we only compare against this approach on MovieLens-1M.
\end{itemize}

The comparison against other hashing-based methods is omitted, as they are either content-based~\cite{DBLP:conf/kdd/LianLG00C17,DBLP:conf/wsdm/ZhangYHDYL18,DBLP:conf/ijcai/Liu0FNLZ18}, designed for explicit feedback~\cite{DBLP:conf/kdd/ZhangWLTYY18}, or outperformed by our baselines~\cite{DBLP:conf/cvpr/LiuHDL14}.

Finally, our candidate generation and re-ranking based framework \textbf{CIGAR}, which first retrieves $c=200$ candidates from the Multi-Index Hash (MIH) table, and adopts ranking models to re-rank the candidates to obtain final recommendations. By default, CIGAR employs HashRec to learn binary user/item embeddings, and BPR-MF\textsuperscript{+} as the re-ranking model. The effect of CIGAR with different candidate generation methods, re-ranking models, and hyper-parameters are also studied in the experiments.
More details on hyperparameter tuning are included in the appendix.

\subsection{Recommendation Performance}

Table~\ref{tab:recommendation} shows Top-N recommendation accuracy on all datasets. Columns (a-1) to (a-4) contain `efficient' recommendation methods (i.e., based on popularity or hashing), while (b-1) to (b-3) represent real-valued ranking models. For hashing-based methods, we use HR@200 to evaluate the performance of candidate generation (i.e., whether the desired item appears in the 200 candidates). 

Not surprisingly, there is a clear gap between hashing-based methods and real-valued methods in terms of HR@10, which confirms that using binary embeddings alone makes it difficult to identify the fine-grained Top-10 ranking due to the compactness of the binary representations. However, we find that the HR@200 of HashRec (and DCF) is significantly higher than the HR@10 of (b-1) to (b-3), which suggests the potential of using hashing-based methods to generate coarse-grained candidates, 
as the HR@200 during the candidate generation stage is 
an upper bound for the Top-10 performance (e.g.~if we have a perfect re-ranking model, the HR@10 would be equal to the HR@200) using the CIGAR framework.

Table~\ref{tab:recommendation} shows that HashRec significantly outperforms hashing-based baselines, presumably due to the $\tanh(\cdot)$ approximation and the use of the advanced \emph{Adam} optimizer.
Hence, we choose HashRec as the candidate generation method in CIGAR by default.
Finally, we see that CIGAR (with HashRec and BPR-MF\textsuperscript{+}) outperforms state-of-the-art recommendation approaches. Note that CIGAR only ranks 200 candidates (generated by HashRec), while BPR-MF, CML and NeuMF rank all  items to obtain the Top-10 results. This suggests that only considering a small number of high-quality candidates is sufficient to achieve satisfactory performance.

\textbf{Comparison against DPR} We perform a comparison with DPR~\cite{DBLP:conf/aaai/ZhangLY17} on the smaller dataset MovieLens-1M as the DPR is hard to scale to other datasets we considered. Following the same data filtering, partitioning, and evaluation scheme, DPR achives an HR@10 of 8.9\%, HR@200 of 55.7\%. In comparison, HashRec's HR@10 is 13.5\%, and HR@200 is 64.6\%. This shows that HashRec outperforms DPR which is also designed for implicit recommendation.

\subsection{Effect of the Number of Candidates}

Figure~\ref{fig:NC} shows the HR@10 of various approaches with different numbers of candidates. We can observe the effect of different candidate generation methods by comparing HashRec, DCF, POP, and RAND with a fixed ranking approach (BPR-MF). CIGAR0 (HashRec+BPR-MF) clearly outperforms alternate approaches, and achieves satisfactory performance (similar to All Items+BPR-MF) on the first three datasets. For larger datasets, more bits and more candidates might be helpful to boost the performance (see 
sec.~\ref{sec:HP}).

However, CIGAR0 merely approximates the performance of All Items+BPR-MF. As we pointed out in section~\ref{sec:COS}, this approach is suboptimal as the vanilla BPR-MF is trained to rank all items, whereas we need to rank a small number of `hard' candidates in the CIGAR framework. By adopting the candidate-oriented sampling strategy to train a BPR-MF model focusing on ranking candidates, we see that CIGAR achieves significant improvements over All Items+BPR-MF. This confirms that the proposed candidate-oriented re-ranking strategy is crucial in helping CIGAR to achieve better performance than the original ranking model.

Note that CIGAR is trained to re-rank the $c$=200 candidates. This may cause the performance drop when ranking more candidates on small datasets like \emph{ML-20M} and \emph{Yelp}.

\subsection{Effects of Candidate-oriented Re-ranking}

In previous sections, we have shown the performance of CIGAR with different candidate generation methods (e.g.~HashRec, DCF, POP). Since CIGAR is a general framework, in this section we examine the performance of CIGAR using CML and NeuMF as the re-ranking model (BPR-MF is omitted here, as results are included in Figure~\ref{fig:NC}), 
so as to investigate whether the candidate-oriented sampling strategy is helpful in general.

Table~\ref{tab:sampling} lists the performance (HR@10) of CIGAR using CML and NeuMF as its ranking model. Due to the high quality of the 200 candidates, HashRec+CML and HashRec+NeuMF can achieve comparable performance compared to rank all items with the same model. Moreover, we can consistently boost the perfomance via re-training the model with the candidate-oriented sampling strategy (i.e., CML\textsuperscript{+} and NeuMF\textsuperscript{+}), which shows the mixed sampling is the key factor for outperforming the vanilla models with exhaustive searches (refer to section~\ref{sec:HP} for more analysis).

\begin{table}[htb]
\centering\small
\caption{Effects of the candidate-oriented re-ranking sampling with different ranking models. $\uparrow$ indicates  better performance than ranking all items with the same model.}\label{tab:sampling}
\vspace{-0.2cm}
\begin{tabularx}{\linewidth}{Xlllll}
\toprule
Approach                                    &ML-20M & Yelp  &Amazon &Goodreads\\  \midrule

All items + CML                             & 21.41 & 2.12  & 3.47  & 3.20\\ 
HashRec + CML                               & 21.27 & 2.33$\uparrow$  & 3.34  & 2.90\\
HashRec + CML\textsuperscript{+}            & 23.62$\uparrow$ & 3.19$\uparrow$  & 4.22$\uparrow$  & 3.31$\uparrow$\\[1mm]
All items + NeuMF                           & 18.43 & 1.70  & 1.59 & 2.20\\
HashRec + NeuMF                             & 18.29 & 2.17$\uparrow$  & 2.70$\uparrow$ & 1.96\\
HashRec + NeuMF\textsuperscript{+}          & 20.83$\uparrow$ & 2.37$\uparrow$ & 2.78$\uparrow$ & 2.74$\uparrow$\\
\bottomrule
\end{tabularx}
\vspace{-0.3cm}
\end{table}

\subsection{Recommendation Efficiency}\label{exp:time}

Efficiently retrieving the Top-N recommended items for users 
is important in real-world, interactive recommender systems.
In Table~\ref{tab:time}, we compare CIGAR with alternative retrieval approaches for different ranking models. For all ranking models, a linear scan can be adopted to retrieve the top-10 items. BPR-MF is based on inner products, hence we adopt the MIP~(Maximum Innder Product) Tree~\cite{DBLP:conf/kdd/RamP12} to accelerate search speed. As CML requires a nearest neighbor search in a Euclidean space, we employ the classic KD-Tree and Ball Tree for retrieval. Since NeuMF utilizes neural networks to estimate preference scores,
we use a GPU to accelerate the scan.

\begin{table}[h]
\centering\small
\caption{Running times for recommending the Top-10 items to 1,000 users.\label{tab:time}}
\vspace{-0.2cm}
\setlength{\tabcolsep}{4pt}
\begin{tabular}{llrrrr}
\toprule
\multirow{2}{*}{Model} & \multirow{2}{*}{\begin{tabular}[c]{@{}c@{}}Retrieval\\Approach \end{tabular}} & \multicolumn{4}{c}{Wall Clock Time(s)} \\
&&Yelp&Amazon&Goodreads&Flipkart\\  \midrule
\emph{\#Items}            &                     & 0.1M   &   0.4M     & 1.6M       &    2.9M    \\
\hline
\multirow{3}{*}{BPR-MF}   & Linear Scan         &   109.0 &   375.7   &  1623.6   & 3076.4 \\
                          & MIP Tree~\cite{DBLP:conf/kdd/RamP12}            &   52.8  &   559.5   & 26.5      & 300.8\\
                          & CIGAR               &   1.2   &   1.5     & 1.6       &   1.9 \\ \hdashline
\multirow{4}{*}{CML}      & Linear Scan         &   375.8 &  1439.9   & 5972.5    & 12367.1 \\
                          & KD Tree             &   18.2  &  40.4    & 162.2     & 169.1  \\
                          & Ball Tree           &   15.4  &  46.3    & 210.7     & 227.8  \\                          
                          & CIGAR               &   1.7   &  2.0     & 2.1       & 2.3\\ \hdashline
\multirow{2}{*}{NeuMF}    & Parallel Scan (GPU)                 &   21.4  &  76.1    & 332.4       & -\\
                          & CIGAR               &   1.8   &  2.1     & 2.3          &-\\     
\bottomrule
\end{tabular}
\setlength{\tabcolsep}{6pt}
\vspace{-0.1cm}
\end{table}

On the largest dataset (Flipkart), CIGAR is at least 1,000 times faster than linear scan for all models, and around 100 times faster than tree-based methods. 
Furthermore,
compared to other methods
the retrieval time of CIGAR increases very slowly with the number of items. Taking CML as an example, from Yelp to Flipkart, the query time for linear scan, KD Tree, and CIGAR increases by around 30x, 9x, and 1.4x (respectively). The fast retrieval speed of CIGAR is mainly due to the 
efficiency of
hashtable lookup, and the small number of high-quality candidates for re-ranking.

Unlike KD-Trees, which are specifically designed for accelerating search in models based in Euclidean spaces, CIGAR is a model-agnostic approach that can efficiently accelerate the retrieval time of almost any ranking model, including neural models.
Meanwhile, as shown previously, CIGAR can achieve better accuracy compared with models that rank all items. We note that MIP Tree performs extremely well on Goodreads. One possible reason is that a few learned vectors have large length\footnote{As shown in the appendix, the regularization coefficient is set to 0 for Goodreads, which may cause such a phenomenon.}, and hence the MIP Tree can quickly rule out most items.

Our approach is also efficient for training, for example, the whole training process of HashRec + BPR-MF\textsuperscript{+} on the largest public dataset \emph{Goodreads} can be finished in 3 hours (CPU only).

\subsection{Hyper-parameter Study}\label{sec:HP}

\begin{figure}[htb]
\vspace{-0.1cm}
\centering\small
\begin{subfigure}[b]{\linewidth}
\centering
\includegraphics[width=.99\textwidth]{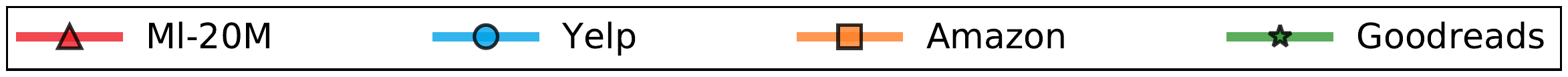}
\end{subfigure}
\begin{subfigure}[b]{0.23\textwidth}
\includegraphics[width=\linewidth]{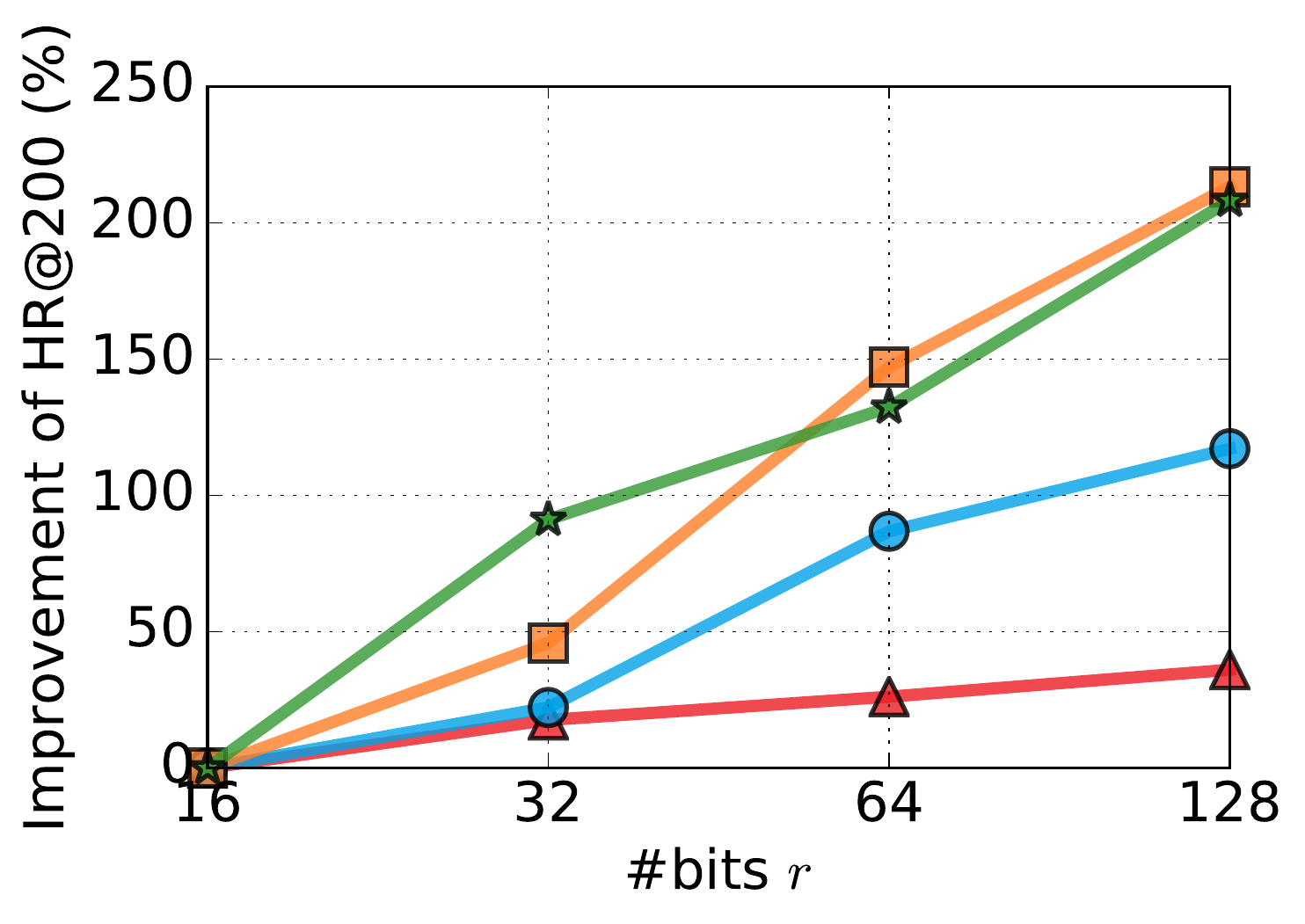}
\subcaption{\#bits $r$}
\end{subfigure}
\begin{subfigure}[b]{0.23\textwidth}
\includegraphics[width=\linewidth]{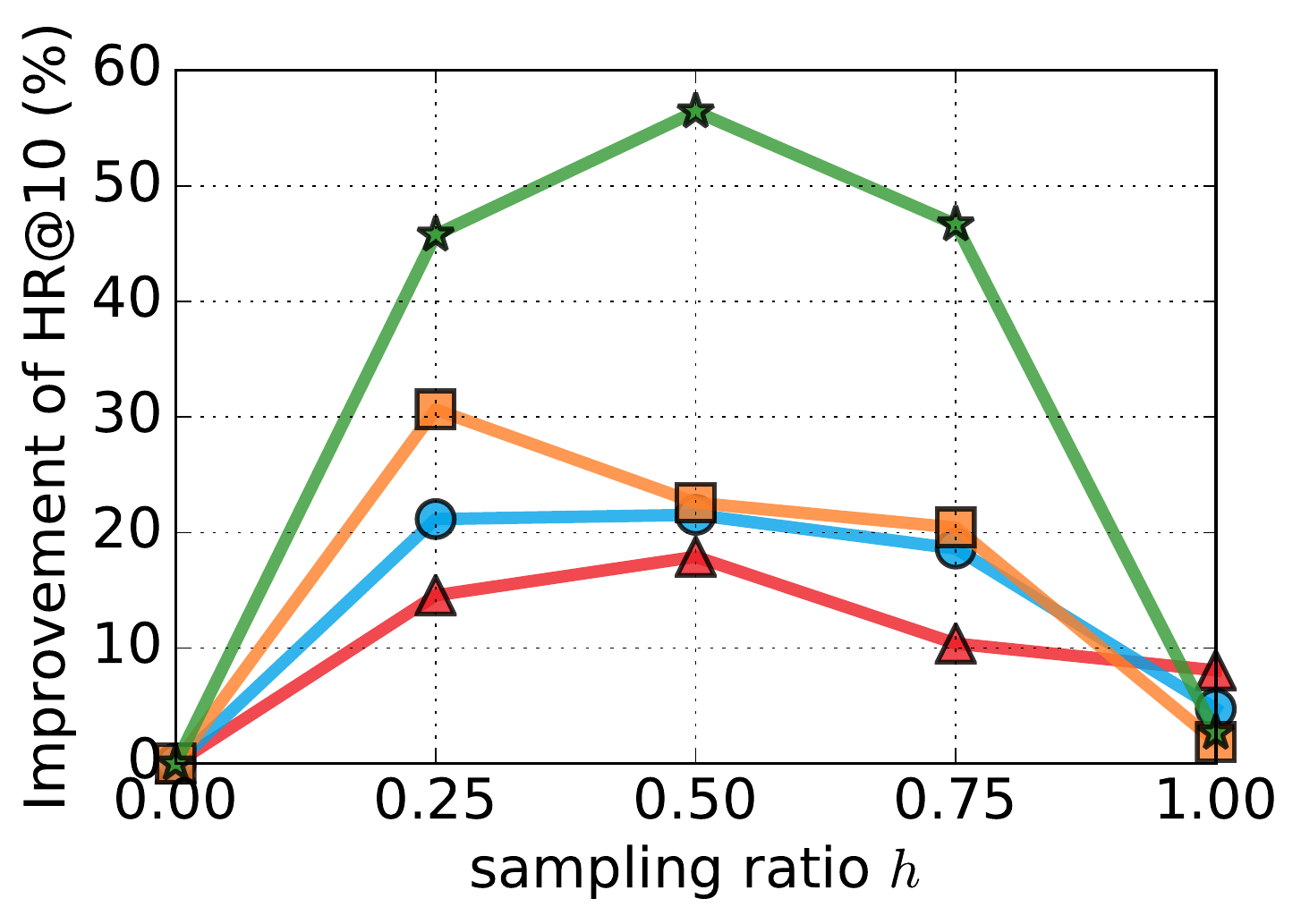}
\subcaption{Sampling ratio $h$}
\end{subfigure}
\vspace{-0.3cm}
\caption{Effects of the Hyper-parameters. (a) improvement of HashRec with more than 16 bits; (b) performance with different sampling ratios.
}
\vspace{-0.3cm}
\label{fig:HP}
\end{figure}

In Figure~\ref{fig:HP}, 
we show the effects of two important hyper-parameters: the number of bits $r$ used in HashRec and the sample ratio $h$ for learning the re-ranking model. For the number of bits, we can clearly observe that more bits leads to better performance. The improvement is generally more significant on larger datasets. For the sample ratio $h$, the best value is around 0.5 for most datasets, thus we choose $h$=0.5 by default. When $h$=1.0, the model seems to overfit due to limited data 
(i.e., a small number of candidates),
and performance 
degrades.
When $h$=0, the model is reduced to the original version which uniformly samples across all items. This again verifies the effect of the proposed candidate-oriented sampling strategy, as it significantly boosts  performance compared to uniform sampling.

\section{Related Work}


\textbf{Efficient Collaborative Filtering} Hashing-based methods have previously been adopted to accelerate retrieval for recommendation. Early methods adopt `two-step' approaches, which first solve a relaxed optimization problem (i.e., binary codes are relaxed to real values), and obtain binary codes by quantization~\cite{DBLP:conf/cvpr/LiuHDL14}. Such approaches often lead to a large quantization error, as learned embeddings in the first stage may be not ideal for quantization. To this end, recent methods jointly optimize the recommendation problem (e.g.~rating prediction) and the quantization loss~\cite{DBLP:conf/sigir/ZhangSLHLC16,DBLP:conf/aaai/ZhangLY17}, leading to better performance compared to two-step methods.
For comparison, we highlight the main differences between our method and existing methods (e.g.~DCF\cite{DBLP:conf/sigir/ZhangSLHLC16} and DPR~\cite{DBLP:conf/aaai/ZhangLY17}) as follows: (1) existing models are often optimized by closed-form solutions involving various matrix operations, whereas HashRec uses a more scalable SGD-based approach; (2) we do not impose bit balance or decorrelation constraints as in DCF and DPR, however in practice we did not observe any performance degradation in terms of accuracy and efficiency; 
(3) we use $\tanh(\beta x)$ to gradually close the gap between binary codes and continuous embeddings during training, which shows effective empirical approximation. To the best of our knowledge, HashRec is the first scalable hashing-based method for implicit recommendation.

Another line of work seeks to accelerate the retrieval process of existing models. For example, approximate nearest neighbor (ANN) search has been heavily studied~\cite{Datar2004LocalitysensitiveHS,DBLP:conf/soda/Yianilos93}, which can be used to accelerate metric-based models like CML~\cite{DBLP:conf/www/HsiehYCLBE17}. Recently, a few approaches have sought to accelerate the maximum inner product search~(MIPS) operation for inner product based models~\cite{DBLP:conf/nips/Shrivastava014,DBLP:conf/kdd/RamP12,DBLP:conf/sigmod/LiCYM17}. However, these approaches are generally model-dependant, and can not easily be adapted to other models. 
For example, it is difficult to accelerate search for models that use complex scoring functions (e.g.~MLP-based models such as NeuMF~\cite{NeuMF}).
In comparison, CIGAR is a model-agnostic approach that can generally expedite search within any ranking model, as we only require the ranking model to scan a short list of candidates.

Inspired by knowledge distillation~\cite{DBLP:journals/corr/HintonVD15}, a recent approach seeks to learn compact models (i.e., with smaller embedding sizes) while maintaining recommendation accuracy~\cite{DBLP:conf/kdd/TangW18}. However, the retrieval complexity is still linear in the number of items.

\textbf{Candidate Generation and Re-ranking}
To build real-time recommender systems, candidate generation has been adopted in industrial settings like \emph{Youtube}'s video recomendation~\cite{DBLP:conf/recsys/DavidsonLLNVGGHLLS10,DBLP:conf/recsys/CovingtonAS16}, \emph{Pinterest}'s related pin recommendation~\cite{DBLP:conf/www/LiuRSKMZLJ17, DBLP:conf/www/EksombatchaiJLL18}, \emph{Linkedin}'s job recommendation~\cite{DBLP:conf/kdd/BorisyukKSZ16}, and \emph{Taobao}'s product recommendation~\cite{DBLP:conf/kdd/ZhuLZLHLG18}. Such industrial approaches often adopt heuristic rules, similarity measurements, and feature engineering specially designed for their own platforms. A closer work to our approach is \emph{Youtube}'s method~\cite{DBLP:conf/recsys/CovingtonAS16} which learns two models for candidate generation and re-ranking. The scoring function in the candidate generation model is the inner product of user and item embeddings, and thus can be accelerated by maximum inner product search via hashing- or tree-based methods. In comparison, our candidate generation method (HashRec) directly learns binary codes for representing preferences as well as building hash tables. To our knowledge, this is the first attempt to adopt hash code learning techniques for candidate generation.

Other than recommender systems, candidate generation has also been adopted in document retrieval~\cite{DBLP:conf/cikm/AsadiL12}, and NLP tasks\cite{,DBLP:conf/emnlp/ZhangWS14}.

\textbf{Learning to Hash} Unlike conventional data structures where hash functions might be designed (or learned) so as to reduce conflicts~\cite{DBLP:conf/focs/DietzfelbingerKMHRT88,DBLP:conf/sigmod/KraskaBCDP18}, we consider similarity-preserving hashing that seeks to map high-dimensional dense vectors to a low-dimensional Hamming space while preserving similarity. Such approaches are often used to accelerate approximate nearest neighbor~(ANN) search and reduce storage cost. A representative example is Locality Sensitive Hashing~(LSH)~\cite{Datar2004LocalitysensitiveHS} which uses random projections as the hash functions. A few seminal works~\cite{DBLP:journals/ijar/SalakhutdinovH09, DBLP:conf/nips/WeissTF08} propose to learn hash functions from data, which is generally more effective than LSH. Recent work focuses on improving the performance, (e.g.) by using better quantization strategies, and adopting DNNs as hash functions~\cite{DBLP:journals/pieee/WangLKC16, DBLP:journals/pami/WangZSSS18}. Such approaches have been adopted for fast retrieval of various content including images~\cite{DBLP:conf/ijcai/LiWK16,DBLP:conf/cvpr/ShenSLS15}, documents~\cite{DBLP:conf/acl/LiLJ14}, videos~\cite{DBLP:journals/tmm/LiongLTZ17}, and products (e.g.~DCF~\cite{DBLP:conf/sigir/ZhangSLHLC16} and HashRec). HashRec directly learns binary codes for users and items, which is essentially a  hash function projecting one-hot vectors into the Hamming Space. We plan to learn hash functions (e.g.~based on DNNs) to map user and item features to binary codes as future work, such that we can adapt to new users and items, which may alleviate cold-start problems.

\section{Conclusion}


We presented new techniques for candidate generation, a critical (but somewhat overlooked) subroutine in recommender systems that seek to efficiently generate Top-N recommendations.
We sought to bridge the gap between two existing modalities of research: methods that advance the state-of-the-art for Top-N recommendation, but are generally inefficient when trying to produce a final ranking; and methods based on binary codes, which are efficient in both time and space but fall short of the state-of-the-art in terms of ranking performance.
In this paper, we developed a new method based on binary codes to handle the candidate generation step, allowing existing state-of-the-art recommendation modules to be adopted to refine the results. A second contribution was to show that performance can further be improved by adapting these modules to be aware of the generated candidates at training time, using a simple weighted sampling scheme.
We showed experiments on several large-scale datasets, where we observed orders-of-magnitude improvements in ranking efficiency, while maintaining or improving upon state-of-the-art accuracy.
Ultimately, this means that we have a general-purpose framework that can improve the scalability of existing recommender systems at test time, and surprisingly does \emph{not} require that we trade off accuracy for speed.

\noindent\textbf{Acknowledgements.} This work is partly supported by NSF \#1750063. We thank Surender Kumar and Lucky Dhakad for their help preparing the Flipkart dataset.



\bibliographystyle{ACM-Reference-Format}
\bibliography{acmart}

\renewcommand{\thesection}{\texorpdfstring{\MakeUppercase{\alph{section}}}{}}
\setcounter{section}{0}

\newpage

\noindent\textbf{\huge{Appendix}}

\setlength{\aboverulesep}{0pt}
\setlength{\belowrulesep}{0pt}
\begin{table*}\small
\caption{Hyper-parameters}
\label{tab:hp}
\begin{tabular}{c|ccc|ccc|cc|cc|cc|cc}
\toprule
\multirow{2}{*}{} & \multicolumn{3}{c|}{DCF}   & \multicolumn{3}{c|}{HashRec}                    & \multicolumn{2}{c|}{BPR-MF}                 & \multicolumn{2}{c|}{BPR-MF\textsuperscript{+}}   & \multicolumn{2}{c|}{CML}  & \multicolumn{2}{c}{NeuMF}             \\
                  & $r$              & $\alpha$ & $\beta$ &  $r$              & $\lambda$ & $m$ & $k$               & $\lambda$ &  $k$                                & $\lambda$  & $k$ & margin & MLP Arch.              &         $k$      \\ \hline
Ml-20M              & \multirow{5}{*}{64} & 0.001  & 0.001  & \multirow{5}{*}{64} & 0.001                       &  16 & \multirow{5}{*}{50} &      0.0001                & \multirow{5}{*}{50}   &      0.01 & \multirow{5}{*}{50} &     0.5 & \multirow{5}{*}{[200,100,50,25]} & \multirow{5}{*}{25} \\
Yelp                &           &   0.0001  &   0.0001  &           &   0.1     &    8     &          &   0.0001  &           &    0.1   &           &    1.0     &           & \\
Amazon              &           &     0.001 &    0.001  &           &     1.0   &     4     &          &    0.01   &           &     0.1   &           &    2.0    &           & \\
Goodreads           &           &     0.001 &    0.001  &           &    0.01   &     4     &          &    0.0    &           &     0.0001   &           &    1.0 &           & \\
Flipkart                &           &     - &    -  &           &   1.0     &     4     &          &     0.001 &           &    0.01    &           &    2.0   &           & \\
\bottomrule
\end{tabular}
\end{table*}

\section{Implementation Details}

We implemented HashRec, BPR-MF, CML, and NeuMF in Tensorflow (version 1.12). For DCF and DPR, we use the MATLAB implementation from the corresponding authors.\footnote{Code for DCF is available online: \url{https://github.com/hanwangzhang/Discrete-Collaborative-Filtering}}

For HashRec, BPR-MF, CML, and NeuMF, we use the \emph{Adam} optimizer with learning rate 0.001 and a batch size of 10000. A multi-processing sampler is used for accelerating data sampling. All models are trained for a maximum of 100 epochs. We evaluate the validation performance\footnote{HR@200 for hashing-based methods, and HR@10 for ranking models.} every 10 epochs, and terminate the training if it doesn't improve after 20 epochs. The bit length $r$ for all hashing-based methods is set to 64, and the embedding size $k$ for ranking models is set to 50.\footnote{we searched among \{10,30,50\} and found that 50 works best for all models on all datasets.}

We use a validation set to search for the best hyper-parameters. The $\ell_2$ regularizer $\lambda$ in HashRec, BPR-MF, and BPR-MF\textsuperscript{+} is selected from \{1, 0.1, 0.01, 0.001, 0.0001, 0\}. For DCF, the user and item regularizers $\alpha$ and $\beta$ are selected from \{0.01,0.001,0.0001\}, and we set $\alpha=\beta$ for fair comparison as other methods use only one regularizer for both users and items. For CML, the norm of metric embeddings is set to 1 following the paper~\cite{DBLP:conf/www/HsiehYCLBE17}, and the margin in the hinge loss is selected from \{0.1, 0.5, 1.0, 2.0\}. For NeuMF, we follow the default configurations~\cite{NeuMF} and a 3-layer pyramid MLP architecture is adopted. As NeuMF has two embeddings for GMF and MLP, we set the embedding size to 25 for each. For DPR, we use the default setting on MovieLens-1M. DPR training on MovieLens-20M did not terminate in 24 hours, hence we only compare against it on MovieLens-1M.

For CIGAR, on all datasets, the number of candidates $c$=200, the scaling factor $\alpha$=$10/r$, sampling ratio $h$=0.5, and $\beta$ is increased as shown in Algorithm~\ref{algo:hasrec}. For multi-index hashing, the number of sub-tables $m$ is set to \{16,8,4\} depending on dataset sizes.

Table~\ref{tab:hp} shows the hyper-parameters we used for each model on all datasets.

\section{Efficiency Test}

We performed the efficiency test (i.e., Section~\ref{exp:time}) on a workstation with a quad-core Intel i7-6700 CPU and a GTX-1080Ti GPU. The GPU is not used except for NeuMF. For MIP Tree~\cite{DBLP:conf/kdd/RamP12}, we use the implementation from the authors,\footnote{\url{https://github.com/gamboviol/miptree}} and the leaf node size is set to 20 following the default setting. For KD-Trees and Ball Trees, we adopt the implementation from the scikit-learn library.\footnote{\url{https://scikit-learn.org/stable/modules/neighbors.html}} A priority queue is employed for choosing the top-k items in linear scan, which costs $O(|\mathcal{I}|\log k)$. The query time of CIGAR consists of obtaining 200 candidates from MIH, performing linear scan on the candidates, and choosing the top-10 items. We assume the queries are independent, and process them sequentially.


\section{Multi-Index Hashing}

We show the procedure of building multi-index hash tables in Algorithm~\ref{algo:mih1}. For search, we gradually increase the search radius $l$ until we obtain enough candidates or reach the maximum radius $l_\text{max}$. Larger radii may retrieve more accurate neighbors but would cost more time, and we set $l_\text{max}$=1 in our experiments. The search procedure is shown in Algorithm~\ref{algo:mih2}.
\begin{algorithm}
\small
\caption{Building MIH}
\label{algo:mih1}
\begin{algorithmic}
\STATE {\textbf{Input:} item binary codes $\d_i\in\{0,1\}^{r}\ i\in\mathcal{I}$, the number of substrings $m$}
\STATE {Initialize $m$ hash tables $H_1$, $H_2$,\dots,$H_m$, each containing $2^{r/m}$ buckets}
\FOR{$\mathit{i} = 1 \to |\mathcal{I}|$}
  \STATE {Split $\d_i$ into $m$ substrings ($s_1$, $s_2$,\ldots,$s_m$)}
  \FOR{$\mathit{j} = 1 \to \mathit{m}$}
      \STATE{Insert item $\mathit{i}$ into the bucket $s_j$ in $H_j$}
  \ENDFOR
\ENDFOR
\STATE {\textbf{Output:} Hash Tables $H$=($H_1$, $H_2$,\dots,$H_m$)}
\end{algorithmic}
\end{algorithm}

\vspace{-0.5cm}

\begin{algorithm}
\small
\caption{Querying in MIH}
\label{algo:mih2}
\begin{algorithmic}
\STATE {\textbf{Input:} Hash Tables $H$=($H_1$, $H_2$,\dots,$H_m$), query codes $\b_u\in\{0,1\}^r$, maximum radius $\mathit{l_{max}}$}, number of candidates $c$
\STATE {$\mathit{S}\gets \emptyset$}
\STATE {Split $\b_u$ into substrings ($s_1$, $s_2$,\ldots,$s_m$)}
\FOR{$\mathit{l} = 0 \to \mathit{l_{max}}$}
  \FOR{$\mathit{j} = 1 \to \mathit{m}$}
    \FOR{bucket $\mathit{b}$ with $d_H(b,s_j)=l$}
      \STATE{$\mathit{S}\gets\mathit{S}\bigcup$ $\{$items in bucket $\mathit{b}$ of table $H_j\}$ }
    \ENDFOR
  \ENDFOR
  \IF{$|\mathit{S}|\geq c$}
    \STATE{\textbf{break}}
  \ENDIF
\ENDFOR
\IF {$|\mathit{S}|\leq c$}
  \STATE{\textbf{return} $\mathit{S}$}
\ELSE
  \STATE{sort items in $\mathit{S}$ according to their Hamming distances to $\b_u$, and form a set $\mathit{S'}$ with the $c$ nearest items}
  \STATE{\textbf{return} $\mathit{S'}$}
\ENDIF
\end{algorithmic}
\end{algorithm}

\clearpage

\end{document}